\documentclass[letterpaper,titlepage,11pt]{article}
\usepackage{hyperref}

\usepackage{amssymb,amsmath,amsfonts}
\usepackage{epsfig}

\def\clock{{\count0=\time
           \divide\count0 60
           \ifnum\count0<10 0\fi\the\count0
           \multiply\count0 -60 \advance\count0 \time
           :\ifnum\count0<10 0\fi \the\count0
         }}

\newcommand{\timestamp}{{\small\vbox{\hbox{\tt\jobname.tex}
\hbox{\the\day/\the\month/\the\year, \clock}}}}

\newcommand{\CH}{\mathcal{H}}

\newcommand{\CO}{\mathcal{O}}

\newcommand{\CN}{\mathcal{N}}

\newcommand{\Z}{\mathbb{Z}}

\newcommand{\R}{\mathbb{R}}

\newcommand{\nn}{\nonumber}

\newcommand{\ds}{\displaystyle}

\newcommand{\tr}{\mathop{{\rm Tr}}}

\newcommand{\vecto}[2]{\left( \begin{array}{c} #1 \\ #2 \end{array} \right) }
\newcommand{\matrto}[4]{\left( \begin{array}{cc} #1 & #2 \\ #3 & #4 \end{array} \right) }

\newcommand{\ads}{\mbox{AdS}}
\newcommand{\gym}{g_{\rm YM}}

\numberwithin{equation}{section}
\begin{document}

\begin{titlepage}

\rightline{\vbox{\small\hbox{\tt hep-th/0611242} }}
\rightline{\vbox{\small\hbox{\tt NORDITA-2006-11-20} }}
 \vskip 1.5 cm

\centerline{\LARGE \bf Magnetic Heisenberg-chain/pp-wave
correspondence} \vskip 1.8cm

\centerline{\large {\bf Troels Harmark$\,^{1}$}, {\bf Kristjan R.\ Kristjansson$\,^{2}$} and {\bf Marta Orselli$\,^{2}$} }

\vskip 0.6cm

\begin{center}
\sl $^1$ The Niels Bohr Institute  \\
\sl  Blegdamsvej 17, 2100 Copenhagen \O , Denmark \\
\vskip 0.4cm
\sl $^2$ Nordita \\
\sl  Blegdamsvej 17, 2100 Copenhagen \O , Denmark \\
\end{center}
\vskip 0.6cm

\centerline{\small\tt harmark@nbi.dk, kristk@nordita.dk, orselli@nbi.dk}

\vskip 1.5cm

\centerline{\bf Abstract} \vskip 0.2cm \noindent We find a
decoupling limit of planar $\CN=4$ super Yang-Mills (SYM) on $\R
\times S^3$ in which it becomes equivalent to the ferromagnetic
$XXX_{1/2}$ Heisenberg spin chain in an external magnetic field. The
decoupling limit generalizes the one found in hep-th/0605234
corresponding to the case with zero magnetic field. The presence of
the magnetic field is seen to break the degeneracy of the vacuum
sector and it has a non-trivial effect on the low energy spectrum.
We find a general connection between the Hagedorn temperature of
planar $\CN=4$ SYM on $\R \times S^3$ in the decoupling limit and
the thermodynamics of the Heisenberg chain. This is used to study
the Hagedorn temperature for small and large value of the effective
coupling. We consider the dual decoupling limit of type IIB strings
on $\ads_5 \times S^5$. We find a Penrose limit compatible with the
decoupling limit that gives a magnetic pp-wave background. The
breaking of the symmetry by the magnetic field on the gauge theory
side is seen to have a geometric counterpart in the derivation of
the Penrose limit. We take the decoupling limit of the pp-wave
spectrum and succesfully match the resulting spectrum to the low
energy spectrum on the gauge theory side. This enables us to match
the Hagedorn temperature of the pp-wave to the Hagedorn temperature
of the gauge theory for large effective coupling. This generalizes
the results of hep-th/0608115 to the case of non-zero magnetic
field.
%\vskip 0.5cm \leftline{\timestamp}

\end{titlepage}

\pagestyle{plain} \setcounter{page}{1}

\tableofcontents

%%%%%%%%%%%%%%%%%%%%%%%%%%%%%%%%%%%%%%%%%%%%%%%%%%%%%%%%%%%%%%

\section{Introduction}
\label{sec:intro}

The AdS/CFT correspondence conjectures that $\CN = 4$ $SU(N)$
super Yang-Mills (SYM) on $\R \times S^3$ is equivalent to type
IIB string theory on $\ads_5\times S^5$
\cite{Maldacena:1997re,Gubser:1998bc,Witten:1998qj}. In
\cite{Harmark:2006di,Harmark:2006ta} a new decoupling limit of
AdS/CFT was introduced. The limit is most naturally expressed as a
limit of the thermal partition functions on both sides of the
correspondence. On the gauge theory side, the limit is
\cite{Harmark:2006di}
\begin{equation}
\label{gaugelim} \Omega \rightarrow 1, \quad \tilde{T} \equiv
\frac{T}{1-\Omega} \ \mbox{fixed}, \quad \tilde{\lambda} \equiv
\frac{\lambda}{1-\Omega} \ \mbox{fixed}, \quad N\ \mbox{fixed},
\end{equation}
where $T$ is the temperature, $\Omega=\Omega_1=\Omega_2$,
$\Omega_3=0$ with $\Omega_i$ being the chemical potentials
corresponding to the three R-charges $J_i$, $i=1,2,3$, for the
$SU(4)$ R-symmetry of $\CN=4$ SYM. Also, $\lambda$ is the 't Hooft
coupling and $N$ is the rank of the gauge group. In the limit
\eqref{gaugelim} all states except for the ones in the $SU(2)$
sector decouple. In fact, the full partition function of $\CN=4$ SYM
on $\R \times S^3$ reduces to ($\tilde{\beta}=1/\tilde{T}$)
\begin{equation}
\label{theZ} Z(\tilde{\beta}) = \tr \left( e^{-\tilde{\beta} (D_0
+ \tilde{\lambda} D_2 )} \right)
\end{equation}
where the trace is over the $SU(2)$ sector only, and the $D_0$ and
$D_2$ operators come from the weak coupling expansion of the
dilatation operator $D = D_0 + \lambda D_2 + \CO(\lambda^{3/2})$,
with $D_0$ being the bare term and $D_2$ the one-loop term. For
planar $\CN=4$ SYM the $\tilde{\lambda} D_2$ term for single-traces
of a fixed length corresponds to the Hamiltonian of the
ferromagnetic $XXX_{1/2}$ Heisenberg spin chain with zero magnetic
field. Thus, weakly coupled planar $\CN=4$ SYM in the limit
\eqref{gaugelim} is equivalent to the Heisenberg chain.

On the dual string theory side the limit is instead in terms of
the angular velocities $\Omega_i$ on the five-sphere, the string
tension $T_{\rm str}$ and the string coupling $g_s$. It takes the
form \cite{Harmark:2006ta}%
\footnote{As we also emphasize in the main text, the decoupling
limit is more useful on the string side when written in the
microcanonical ensemble. See \cite{Harmark:2006ta} for the
microcanonical version of the decoupling limit \eqref{stringlim}.}
\begin{equation}
\label{stringlim} \Omega \rightarrow 1, \quad \tilde{T} \equiv
\frac{T}{1-\Omega} \ \mbox{fixed}, \quad \tilde{T}_{\rm str} \equiv
\frac{T_{\rm str}}{\sqrt{1-\Omega}} \ \mbox{fixed}, \quad
\tilde{g}_s \equiv \frac{g_s}{1-\Omega} \ \mbox{fixed},
\end{equation}
again with $\Omega=\Omega_1=\Omega_2$ and $\Omega_3=0$. We see
that the limit \eqref{stringlim} involves taking the string
tension $T_{\rm str}$ and the string coupling $g_s$ to zero.
Therefore, the decoupling limit of AdS/CFT found in
\cite{Harmark:2006di,Harmark:2006ta} leads to a correspondence
between a decoupled sector of weakly coupled $\CN=4$ SYM and
weakly coupled string theory.

If we in particular consider planar $\CN =4$ SYM then this is dual
to free type IIB string theory on $\ads_5\times S^5$, and the
decoupling limit gives us therefore in this case a triality between
the Heisenberg chain, a limit of weakly coupled $\CN =4$ SYM and a
zero tension limit of free string theory on $\ads_5\times S^5$. In
\cite{Harmark:2006ta} this was tested for large $\tilde{\lambda}$
and large $J=J_1+J_2$, with the successful result that the spectra
of the gauge theory and string theory sides match. On the gauge
theory side the spectrum corresponds to the spectrum of magnons in
the Heisenberg chain. On the string theory side the spectrum is
obtained from taking the decoupling limit of the spectrum for a
particular pp-wave background. The matching of the spectra
furthermore leads to the result that the Hagedorn temperature, as
computed on the gauge theory/spin chain side, matches with the
Hagedorn temperature computed on the string theory side for the
pp-wave background, for large $\tilde{\lambda}$
\cite{Harmark:2006ta}.

That the Hagedorn temperature on the gauge theory and string theory
sides are dual to each other has been proposed in
\cite{Witten:1998zw,Sundborg:1999ue,Polyakov:2001af,Aharony:2003sx}.
The result of \cite{Harmark:2006ta} thus finds a region of AdS/CFT
where it can be checked explicitly that the Hagedorn temperatures
indeed match.

In this paper we consider a modification of the decoupling limit
\eqref{gaugelim} for $\CN=4$ SYM on $\R \times S^3$ such that it
becomes equivalent to the ferromagnetic Heisenberg $XXX_{1/2}$ spin
chain in an external magnetic field. We have again $\Omega_3=0$ and
we define $\Omega = (\Omega_1+\Omega_2)/2$ and $h =
(\Omega_1-\Omega_2)/2$. The new decoupling limit is then
\begin{equation}
\label{gaugelimmagn} \Omega \rightarrow 1, \quad \tilde{T} \equiv
\frac{T}{1-\Omega} \ \mbox{fixed} , \quad \tilde{h} \equiv
\frac{h}{1-\Omega} \ \mbox{fixed} , \quad \tilde{\lambda} \equiv
\frac{\lambda}{1-\Omega} \ \mbox{fixed}, \quad N\ \mbox{fixed}.
\end{equation}
This limit reduces to \eqref{gaugelim} when $\tilde{h}=0$. The
full partition function of $\CN=4$ SYM on $\R \times S^3$ now
reduces to
\begin{equation}
\label{theZmagn} Z(\tilde{\beta},\tilde{h}) = \tr \left(
e^{-\tilde{\beta} (D_0 + \tilde{\lambda} D_2 - 2 \tilde{h} S_z)}
\right)
\end{equation}
where the trace is, as before, over the $SU(2)$ sector. For planar
$\CN=4$ SYM the $\tilde{\lambda} D_2 - 2\tilde{h} S_z$ part of the
Hamiltonian corresponds to the Hamiltonian of the ferromagnetic
$XXX_{1/2}$ Heisenberg spin chain with a magnetic field of magnitude
$2\tilde{h}$.

For the zero magnetic field case \eqref{theZ} the Heisenberg chain
$\tilde{\lambda} D_2$ has a degenerate vacuum sector. In fact there
is a vacuum state for each value of the total spin $S_z$. The
introduction of the magnetic field in \eqref{theZmagn} gives the
interesting effect that the degeneracy is broken and only a single
vacuum remains. As we explain in the main text, this is fundamental
to the understanding of the physics of $\CN=4$ SYM in the modified
limit \eqref{gaugelimmagn}. In particular, it is responsible for a
non-trivial modification of the spectrum for large $\tilde{\lambda}$
and $J$, and it also gives a non-trivial effect for the Hagedorn
temperature.

On the string theory side we obtain again the spectrum from a
decoupling limit of the spectrum of a particular pp-wave background.
For the zero magnetic field this pp-wave background is the maximally
supersymmetric pp-wave background \cite{Blau:2001ne}, however, not
in the coordinate system used in the gauge-theory/pp-wave
correspondence of BMN \cite{Berenstein:2002jq}, but instead in a
coordinate system where the pp-wave background has a flat direction,
i.e.\ an explicit isometry \cite{Michelson:2002wa,Bertolini:2002nr}.
This flat direction corresponds to the degenerate vacuum sector on
the gauge theory side \cite{Harmark:2006ta}. The flat direction
pp-wave background can be seen as the BMN pp-wave background rotated
with constant angular velocity in a plane with the critical velocity
for which the quadratic terms disappear. With the magnetic field,
the pp-wave background instead corresponds to rotating with a
constant angular velocity that is near the critical angular
velocity.

Since we are not at the critical angular velocity the explicit
isometry of the flat direction is broken. This is the string dual
version of the breaking of the degeneracy of the vacuum sector on
the gauge theory side caused by the magnetic field. We observe that
for this reason the decoupled sector of the near-critical pp-wave
used in this paper resembles much more the pp-wave background used
by BMN than the pp-wave with a flat direction.

The pp-wave background with a near-critical angular velocity can
also be seen as a magnetic pp-wave background, in the sense that the
off-diagonal terms in the metric are analogous to a magnetic field.
We therefore dub the background a {\sl magnetic pp-wave background},
so in this sense we can say that we have obtained a correspondence
between the magnetic Heisenberg chain and a magnetic pp-wave
background.

We match successfully both the spectrum and the Hagedorn temperature
as found from the gauge-theory/spin-chain side and from the string
theory side. This provides a new example of a direct correspondence
between a sector of weakly coupled gauge theory and free string
theory which can be seen as an extension of that of
Ref.~\cite{Harmark:2006ta}.

%%%%%%%%%%%%%%%%%%%%%%%%%%%%%%%%%%%%%%%%%%%%%%%%%%%%%%%%%%%%%%
\section{Gauge theory side: The magnetic Heisenberg chain}
\label{sec:spinchain}

In this section we introduce a new decoupling limit of thermal $\CN
= 4$ super Yang-Mills (SYM) on $\R \times S^3$ in which $\CN=4$ SYM
reduces to a quantum mechanical theory on the $SU(2)$ sector. This
limit can be seen as a generalization of the $SU(2)$ decoupling
limit found in \cite{Harmark:2006di}. We show that in the decoupling
limit, planar $\CN = 4$ SYM becomes equivalent to the ferromagnetic
$XXX_{1/2}$ Heisenberg spin chain with a magnetic field. This should
be contrasted to the $SU(2)$ decoupling limit of
\cite{Harmark:2006di} in which planar $\CN=4$ SYM is equivalent to
the Heisenberg chain without a magnetic field. We use the connection
to the Heisenberg spin chain to compute the Hagedorn temperature for
small and large values of the effective coupling $\tilde{\lambda}$,
and also to compute the spectrum for large $\tilde{\lambda}$.

%%%%%%%%%%%%%%%%%%%%%%%%%%%%%%%%%%%%%%%
\subsection{New decoupling limit}
\label{sec:declim}

As reviewed in the Introduction, the decoupling limit
\eqref{gaugelim} of $\CN=4$ SYM on $\R \times S^3$ with gauge group
$SU(N)$ was found recently
in \cite{Harmark:2006di}.%
\footnote{The decoupling limit can be implemented both for $SU(N)$
and $U(N)$ as the gauge group. Since in this paper we only consider
the planar limit in detail, the difference between the two gauge
groups is not important.} This limit can be expressed in terms of
thermal $\CN = 4$ SYM on $\R \times S^3$ as a limit of the grand
canonical partition function depending on the temperature $T$ and
the three chemical potentials $\Omega_i$, $i=1,2,3$, corresponding
to the three R-charges $J_i$, $i=1,2,3$, for the $SU(4)$ R-symmetry
of $\CN=4$ SYM. In the limit \eqref{gaugelim} the chemical
potentials are chosen such that $\Omega_3=0$ and
$\Omega_1=\Omega_2=\Omega$, thus reducing the four variables in the
grand canonical partition function to $T$ and $\Omega$. Note that
$\lambda$ is the 't Hooft coupling defined for convenience as
\begin{equation}
\lambda = \frac{\gym^2 N}{4\pi^2}
\end{equation}
with $\gym$ being the Yang-Mills coupling and $N$ the rank of the
gauge group.

In the new decoupling limit that we introduce in this paper we still
take $\Omega_3=0$ but $\Omega_1$ and $\Omega_2$ are no longer
required to be equal. They still both go to one in the limit, but in
such a way that the difference $\Omega_1-\Omega_2$ also plays a
non-trivial role. It is therefore natural to define
\begin{align}
\label{omegah} \Omega \equiv \frac{1}{2}(\Omega_1 + \Omega_2),
\qquad h \equiv \frac{1}{2}(\Omega_1 - \Omega_2).
\end{align}
We see that with $h=0$ we have $\Omega_1=\Omega_2=\Omega$ as in
\eqref{gaugelim}. In accordance with this, it is useful to combine
the R-charges $J_1$ and $J_2$ into the following charges
\begin{align}
\label{jsz} J \equiv J_1 + J_2, \qquad S_z \equiv \frac{1}{2}
(J_1-J_2).
\end{align}
With this, we can write the grand canonical partition function as
\begin{equation}
Z(\beta,\Omega,h) \label{eq:grandcan}
= {\tr}_M\exp\left(-\beta D + \beta ( \Omega_1 J_1 + \Omega_2 J_2 ) \right) \\
= {\tr}_M\exp\left(-\beta D + \beta \Omega  J  +  2\beta h
S_z\right).
\end{equation}
The trace is taken over all gauge singlet states, which correspond
to all linear combinations of the multi-trace operators, denoted
here as the set $M$. We write furthermore the inverse temperature as
$\beta = 1/T$. In Eq.~\eqref{eq:grandcan} $D$ is the dilatation
operator which in weakly coupled $\mathcal{N}=4$ SYM can be expanded
in powers of the 't~Hooft coupling as
\cite{Beisert:2003tq,Beisert:2004ry}
\begin{align}
\label{eq:Dexpanded} D = D_0 + \lambda D_2 + \lambda^{3/2} D_3 +
\lambda^2 D_4 + \mathcal{O}(\lambda^{5/2})
\end{align}
where $D_0$ is the bare scaling dimension and $D_2$ is the one-loop
dilatation operator. The partition function can therefore be written
as
\begin{align}
\label{eq:partbflimit} Z(\beta,\Omega,h) = {\tr}_M\exp\left( -\beta
(D_0-J)
 -\beta (1-\Omega) J  -\beta \lambda D_2 +  2\beta h S_z
-\beta\lambda \mathcal{O}(\lambda^{1/2}) \right).
\end{align}

We introduce now the new decoupling limit of thermal $\CN=4$ SYM on
$\R \times S^3$ with gauge group $SU(N)$ given by
\begin{align}
\label{declim} \Omega \to 1, \quad \tilde{T} \equiv
\frac{T}{1-\Omega} \ \mbox{fixed}, \quad \tilde h \equiv
\frac{h}{1-\Omega} \ \mbox{fixed}, \quad \tilde \lambda \equiv
\frac{\lambda}{1-\Omega} \ \mbox{fixed}, \quad N\ \mbox{fixed}.
\end{align}
{}From the first term in the exponent of Eq.~\eqref{eq:partbflimit}
we see that since $\beta \rightarrow \infty$ the states that are not
in the $SU(2)$ sector with $D_0 = J$ have an exceedingly small
weight factor and are therefore decoupled \cite{Harmark:2006di}.
From the last term in the exponent we see that all the terms of the
dilatation operator \eqref{eq:Dexpanded} beyond one loop vanish in
the limit. We can therefore write the partition function as
\begin{align}
\label{decpartfct} Z(\tilde \beta,\tilde h) = {\tr}_\mathcal{H}
\left(e^{-\tilde\beta H}\right)
\end{align}
where $\tilde{\beta}=1/\tilde{T}$, the decoupled Hamiltonian is
given by
\begin{align}
\label{eq:decoupledH} H = D_0 + \tilde \lambda D_2 - 2\tilde h S_z,
\end{align}
and we have restricted the trace to the $SU(2)$ sector
\begin{align}
\mathcal{H} = \left\{ \alpha \in M \big\vert (D_0 - J) \alpha =
0\right\}.
\end{align}
More precisely, the set of operators $\CH$ in the $SU(2)$ sector
consists of all linear combinations of the multi-trace operators
\begin{align}
\label{su2op}  \tr( A^{(1)}_1 A^{(1)}_2 \cdots A^{(1)}_{L_1} )\tr(
A^{(2)}_1 A^{(2)}_2 \cdots A^{(2)}_{L_2} ) \cdots \tr( A^{(k)}_1
A^{(k)}_2 \cdots A^{(k)}_{L_k} ) , \quad A^{(i)}_{j} \in \{ Z,X \}
\end{align}
where the letters $Z$ and $X$ are the two complex scalars of the
gauge theory with R-charge weights $(1,0,0)$ and $(0,1,0)$,
respectively.

Our new decoupling limit \eqref{declim} generalizes the limit
\eqref{gaugelim} found in \cite{Harmark:2006di} since it reduces to
that for $\tilde{h}=0$. In the new decoupling limit \eqref{declim}
we get a decoupled quantum mechanical subsector of $\CN=4$ SYM on
$\R \times S^3$, as in the limit \eqref{gaugelim}. However, we can
now in principle compute the full partition function
\eqref{decpartfct} for any value of $\tilde{\lambda}$, $\tilde{h}$
and $N$. Therefore, we have an extra parameter as compared to the
decoupled quantum mechanical sector arising from the limit
\eqref{gaugelim}. As we shall see below, the extra parameter
$\tilde{h}$ can be regarded both as a magnetic field, and also as an
effective chemical potential.

\subsubsection*{Planar limit and the Heisenberg chain}

We consider now the planar limit $N=\infty$ of $\CN=4$ SYM on $\R
\times S^3$. In this case, we can single out the single-trace
sector, and the full partition function can be found from the
single-trace partition function. The single-trace operators in the
$SU(2)$ sector are built from linear combinations of the following
operators
\begin{align}
\label{singsu2} \tr\left(A_1 A_2 \cdots A_L \right), \quad A_i \in
\{Z,X\}.
\end{align}

Single-trace operators of a fixed length $L$ can be regarded as
states for a spin chain \cite{Minahan:2002ve}. This is done by
interpreting the operator $S_z=(J_1-J_2)/2$ defined in \eqref{jsz}
as the value of the spin for each site of a spin chain. We see that
$Z$ has $S_z=1/2$ while $X$ has $S_z=-1/2$ and hence we regard $Z$
and $X$ as spin up and spin down, respectively. Single-trace
operators are then mapped to states of the spin chain, and $S_z$ for
a single-trace operator becomes the total spin for the corresponding
state of the spin chain.

For a chain of length $L$, the $D_2$ term in \eqref{eq:decoupledH}
may be expressed as \cite{Minahan:2002ve}
\begin{align}
\label{D2} D_2 = \frac{1}{2}\sum_{i=1}^L \left(I_{i,i+1} -
P_{i,i+1}\right)
\end{align}
where $I_{i,i+1}$ is the identity operator and $P_{i,i+1}$ is the
permutation operator acting on letters at position $i$ and $i+1$.
Through the spin chain interpretation, the $D_2$ operator in
\eqref{D2} becomes the Hamiltonian of the ferromagnetic $XXX_{1/2}$
Heisenberg spin chain of length $L$ with zero magnetic field
\cite{Minahan:2002ve}. Using this, we see now that the $\tilde
\lambda D_2 - 2\tilde h S_z$ part of our decoupled Hamiltonian
\eqref{eq:decoupledH} is the Hamiltonian for a ferromagnetic
$XXX_{1/2}$ Heisenberg spin chain of length $L$ with nearest
neighbor coupling $\tilde \lambda$ in an external magnetic field of
magnitude $2\tilde h$ that couples to the spins through a Zeeman
term.

The full partition function of planar $\CN = 4$ SYM on
$\mathbb{R}\times S^3$ in the decoupling limit \eqref{declim} is
therefore \cite{Harmark:2006di}
\begin{align}
\label{eq:logZ} \log Z(\tilde \beta, \tilde h) = \sum_{n=1}^\infty
\sum_{L=1}^\infty \frac{1}{n} e^{-n\tilde\beta
L}Z_L^{(XXX)}(n\tilde\beta,\tilde h)
\end{align}
where
\begin{align}
\label{zxxx} Z_L^{(XXX)}(\tilde\beta, \tilde h) = {\tr}_L \left(
e^{-\tilde\beta H_{XXX}}\right)
\end{align}
is the partition function for the ferromagnetic $XXX_{1/2}$
Heisenberg spin chain of length $L$ with Hamiltonian
\begin{align}
\label{ham} H_{XXX}=\tilde{\lambda} D_2-2\tilde hS_z .
\end{align}

It is important to note that $\tilde{h}$ is bounded as $0 \leq
\tilde{h} \leq 1$. The lower bound comes from the fact that we
choose $\tilde{h}$ to be positive. This choice means that the
ground state (in the single-trace sector) is
\begin{equation}
\label{gstate} \tr ( Z^L ).
\end{equation}
The upper bound comes from the fact that the partition function
\eqref{eq:grandcan} is only well-defined in the planar limit for
$|\Omega_i| \leq 1$, $i=1,2$. We choose $\Omega_i$, $i=1,2$, to be
positive. Assuming $\Omega < 1$ we get from $\Omega_1 = \Omega +
\tilde{h} (1-\Omega)$ that $\Omega_1 \leq 1$ implies $\tilde{h} \leq
1$. Note in particular that the critical value $\Omega_1=1$
corresponds to $\tilde{h}=1$. We comment more below on having
$\tilde{h}$ equal or close to one.

\subsubsection*{Hagedorn temperature from Heisenberg chain}

The partition function of free planar $\CN = 4$ SYM on $\R \times
S^3$ exhibits a singularity at a certain temperature $T_H$
\cite{Sundborg:1999ue,Polyakov:2001af,Aharony:2003sx}. The
temperature $T_H$ is a Hagedorn temperature for planar $\CN=4$ SYM
on $\R \times S^3$ since the density of states goes like $e^{E/T_H}$
for high energies $E\gg 1$ (we work in units with radius of the
$S^3$ set to one). Moreover, for large $N$ the transition at $T_H$
resembles the confinement/deconfinement phase transition in QCD.
Turning on the coupling $\lambda$ and the chemical potentials
$\Omega_i$ the Hagedorn singularity for planar $\CN=4$ SYM on $\R
\times S^3$ persists, at least for $\lambda \ll 1$
\cite{Spradlin:2004pp,Yamada:2006rx,Harmark:2006di}.

In Ref.~\cite{Harmark:2006ta} a precise relation was found between
the Hagedorn temperature of planar $\CN=4$ SYM on $\R \times S^3$ in
the decoupling limit \eqref{gaugelim} and the thermodynamics of the
Heisenberg chain in the thermodynamic limit. We extend now this
relation to the case with non-zero external magnetic field.

Following \cite{Harmark:2006ta}, we define the function
$V(\tilde\beta)$ as
\begin{equation}
\label{defV} V(\tilde\beta) \equiv \lim_{L \rightarrow \infty}
\frac{1}{L} \log \left[ {\tr}_L \left( e^{- \tilde\beta H_{XXX}}
\right) \right]
\end{equation}
where $H_{XXX}$ is given in \eqref{ham}. The function
$V(\tilde\beta)$ is related to the thermodynamic limit of the free
energy per site of the Heisenberg chain by $f =
-V(\tilde\beta)/\tilde\beta$.
As in Ref.~\cite{Harmark:2006ta}, the partition function
\eqref{eq:logZ} reaches a Hagedorn singularity (for $n=1$) if
$\tilde{\beta}$ decreases to the critical value $\tilde{\beta}_H$
given by \cite{Harmark:2006ta}
\begin{equation}
\label{genhag} \tilde{\beta}_H = V (\tilde\beta_H).
\end{equation}
Thus, we have obtained a direct relation between the Hagedorn
temperature of planar $\CN=4$ SYM on $\R \times S^3$ in the
decoupling limit \eqref{declim} and the thermodynamics of the
Heisenberg chain with a magnetic field in the thermodynamic limit.

%%%%%%%%%%%%%%%%%%%%%%%%%%%%%%%%%%%%%%%%%%%%%%%
\subsection{Hagedorn temperature for small $\tilde \lambda$}
\label{sec:smalllamb}

In this section we calculate the Hagedorn temperature for small
$\tilde\lambda$. We describe how one can obtain the Hagedorn
temperature to any desired order in $\tilde{\lambda}$ by using the
relation \eqref{genhag} to the free energy of the Heisenberg chain
with a magnetic field. We consider furthermore in detail the
Hagedorn temperature as function of the magnetic field for
$\tilde{\lambda}=0$. Subsequently, we make a consistency check on
the computation of the Hagedorn temperature to one-loop order by
computing it from the pole of the gauge theory partition function.
That the two methods agree provides a non-trivial check of our
decoupling limit and also shows the power of the Heisenberg chain
description since it gives the same result in a much more efficient
way.

\subsubsection*{Hagedorn temperature from the Heisenberg chain}

Eq.~\eqref{genhag} relates the Hagedorn temperature $\tilde{T}_H =
1/\tilde{\beta}_H$, for a given value of $\tilde{\lambda}$ and
$\tilde{h}$, to the thermodynamics of the Heisenberg chain. We now
demonstrate how powerful this connection is by showing how one can
compute the Hagedorn temperature for $\tilde{\lambda} \ll 1$ to any
desired order in $\tilde{\lambda}$.

The small $\tilde\lambda$ limit corresponds to the high-temperature
limit of the magnetic Heisenberg chain given by $\tilde{\lambda}
\tilde{\beta} \ll 1$ with $\tilde{\beta}\tilde{h}$ fixed. This limit
is well studied in the literature, see e.g.~\cite{takahashi}. The
high-temperature expansion of $V(\tilde{\beta})$ can be obtained to
any desired order in $\tilde{\beta}\tilde{\lambda}$ for fixed
$\tilde\beta\tilde{h}$ using a powerful integral equation technique
derived in \cite{Takahashi:2000,Takahashi:2001} and applied to this purpose in \cite{PhysRevLett.89.117201}. In order to apply this to our case, we
introduce the function $u(x)$ defined by the integral equation
\begin{align}
\label{ueq} u(x) = 2\cosh\left(\tilde\beta\tilde h\right) +
\oint_C \frac{dy}{2\pi i}
\left(\frac{\exp\left(\frac{-2\tilde\beta\tilde\lambda}{y(y+2i)}\right)}{x-y-2i}
+
\frac{\exp\left(\frac{-2\tilde\beta\tilde\lambda}{y(y-2i)}\right)}{x-y+2i}
\right)\frac{1}{u(y)}
\end{align}
where the path $C$ is a counterclockwise loop around the origin.
From the function $u(x)$ one then finds $V(\tilde\beta)$ as
\begin{align}
\label{VVeq} V(\tilde\beta) = \log\left( u(0)\right).
\end{align}
In the high-temperature limit, one first determines $u(x)$ as an
expansion in powers of $\tilde\beta\tilde\lambda$ (for fixed
$\tilde{\beta}\tilde{h}$) order by order from the integral equation
\eqref{ueq}. Plugging the resulting expansion into Eq.~\eqref{VVeq},
one then finds the high-temperature expansion of $V(\tilde{\beta})$.

Having determined the high-temperature expansion of
$V(\tilde{\beta})$, it is simple to get the small $\tilde{\lambda}$
expansion of $\tilde{T}_H$ using \eqref{genhag}. To illustrate this,
consider the first few terms of the expansion of $V(\tilde{\beta})$
in powers of $\tilde{\beta}\tilde{\lambda}$ for fixed
$\tilde{\beta}\tilde{h}$ \cite{PhysRevLett.89.117201}
\begin{align}
\label{specV} V(\tilde\beta) = \log (2\cosh (\tilde\beta\tilde h))
    - \frac{\tilde\beta\tilde\lambda}{4} \left(1-\tanh^2(\tilde\beta\tilde h) \right)
    +\frac{3(\tilde\beta\tilde\lambda)^2}{32} \left(1-\tanh^4(\tilde\beta\tilde h) \right)
    + \mathcal{O}((\tilde\beta\tilde\lambda)^3).
\end{align}
Plugging this expansion into Eq.~\eqref{genhag} we obtain the
Hagedorn temperature to the desired order. Writing
\begin{align}
\label{eq:betaexpanded} \tilde \beta_\textrm{H} = \tilde
\beta_\textrm{H}^{(0)} + \tilde\lambda \tilde
\beta_\textrm{H}^{(1)} + \tilde\lambda^2 \tilde
\beta_\textrm{H}^{(2)} + \cdots
\end{align}
we find using Eq.~\eqref{specV} in Eq.~\eqref{genhag}
\begin{align}
\label{zerohag}
%\label{eq:treelevelhagedornbeta}
\tilde \beta_\textrm{H}^{(0)} &=
\log(2\cosh(\tilde \beta_\textrm{H}^{(0)} \tilde h)), \\
\label{onehag} \tilde \beta_\textrm{H}^{(1)}
    &= -\frac{\tilde\beta^{(0)}_\textrm{H}}{4}
    \frac{1-B^2} {1 - \tilde h B}, \\
\tilde \beta_\textrm{H}^{(2)}
    &= -\tilde\beta^{(0)}_\textrm{H}\frac{1-B^2}{32(1-\tilde h B)^3}
    \bigg(-2 + 3 \tilde\beta^{(0)}_\textrm{H} - \tilde\beta^{(0)}_\textrm{H} \tilde h^2
+ (2 \tilde h - 2 \tilde\beta^{(0)}_\textrm{H} \tilde h) B \nonumber \\
&\quad + (2 + 3 \tilde\beta^{(0)}_\textrm{H} +
\tilde\beta^{(0)}_\textrm{H} \tilde h^2) B^2 + (-2\tilde h - 10
\tilde\beta^{(0)}_\textrm{H}\tilde h) B^3 + 6
\tilde\beta^{(0)}_\textrm{H}\tilde h^2 B^4\bigg),
\end{align}
where we have defined
$B\equiv\tanh(\tilde\beta_\textrm{H}^{(0)}\tilde h)$. Note that
$\tilde\beta^{(0)}_\textrm{H}$ is only given in terms of an implicit
equation, and the other coefficients are then written in terms of
$\tilde\beta^{(0)}_\textrm{H}$. The above expansion of the Hagedorn
temperature to order $\tilde{\lambda}^2$ reduces to the one found in
\cite{Harmark:2006ta} for $\tilde{h}=0$.

In conclusion, we can determine $\tilde{T}_H$ for small
$\tilde{\lambda}$ to any desired order by computing the
high-temperature expansion of the function $u(x)$ from the
integral equation \eqref{ueq}, and then plugging the result into
Eqs.~\eqref{VVeq} and \eqref{genhag}.

\begin{figure}[ht]
\centerline{\epsfig{file=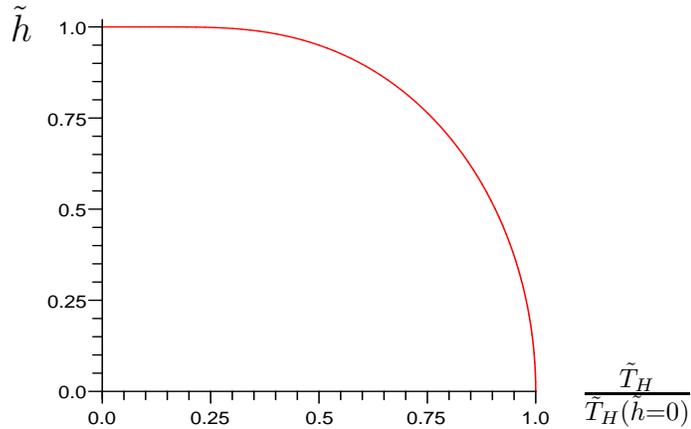,width=8cm,height=7cm} }
\caption{The Hagedorn temperature $\tilde{T}_H$ as function of
$\tilde{h}$ for $\tilde{\lambda} = 0$, with $\tilde{T}_H
({\tilde{h}=0}) = 1/\log 2$.\label{fig:htcurve_weak}}
\begin{picture}(0,0)(0,0)
\put(117,215){\Large $\tilde h$} \put(333,78){\Large
$\frac{\tilde{T}_H}{\tilde{T}_H ({\tilde{h}=0})}$}
\end{picture}
\end{figure}

Consider the zeroth order part of the Hagedorn temperature given by
Eq.~\eqref{zerohag}. From this implicit equation we can find
$\tilde{T}_H$ as a function of $\tilde{h}$. For small $\tilde{h}$ we
have the expansion
\begin{equation}
\tilde{T}_H = \frac{1}{\log 2} - \frac{1}{2} \tilde{h}^2 -
\frac{\log 2}{12} ( 3 - \log 2 ) \tilde{h}^4 + \CO(\tilde{h}^6).
\end{equation}
To understand better the behavior of $\tilde{T}_H$ for $\tilde{h}$
in the full range from 0 to 1 we have solved Eq.~\eqref{zerohag}
numerically and plotted the result in Fig.~\ref{fig:htcurve_weak}.
We see from Fig.~\ref{fig:htcurve_weak} that the Hagedorn
temperature $\tilde{T}_H$ approaches zero for $\tilde{h} \rightarrow
1$. This confirms our upper bound on $\tilde{h}$ stating that
$\tilde{h}\leq 1$, as found in Section \ref{sec:declim}, since
planar $\CN=4$ SYM in the decoupling limit \eqref{declim} is only
well-defined below the $\tilde{T}_H ( \tilde{h} )$ curve in
Fig.~\ref{fig:htcurve_weak}. This has the consequence that we only
reach $\tilde{h}=1$ when $\tilde{T}_H=0$.

Taking into account the corrections in $\tilde{\lambda}$, one can
in principle plot the $\tilde{T}_H ( \tilde{h} )$ curve for small
values of $\tilde{\lambda}$. From the first few corrections
computed above, it is apparent that we still have that
$\tilde{T}_H \rightarrow 0$ as $\tilde{h} \rightarrow 1$.

It is interesting to notice that the $\tilde{T}_H ( \tilde{h} )$
curve in Fig.~\ref{fig:htcurve_weak} has a strong resemblance with
the curves found in \cite{Yamada:2006rx,Harmark:2006di} for $T_H$
as function of the chemical potentials in the full planar $\CN=4$
SYM with zero 't Hooft coupling. Thus it makes sense to regard
$\tilde{h}$ as an effective chemical potential for the decoupled
sector of $\CN=4$ SYM in the limit \eqref{declim}, in the same
sense that $\tilde{T}$ can be regarded as an effective
temperature.

\subsubsection*{Hagedorn temperature from the pole of the partition function}

As a consistency check, we compute here the Hagedorn temperature
$\tilde{T}_H$ to order $\tilde{\lambda}$ directly from the partition
function. This is a check on the consistency of the decoupling limit
\eqref{declim} and of the relation \eqref{genhag} between the
Hagedorn temperature and the thermodynamics of the Heisenberg chain.

The Hagedorn temperature is given by the location of the first
pole of the full partition function \eqref{eq:grandcan} in the
planar limit $N=\infty$. Using this fact, the Hagedorn temperature
can be calculated to first order in $\tilde\lambda$ by the
technique introduced in \cite{Spradlin:2004pp} and extended to the
case with chemical potentials in \cite{Harmark:2006di}.

The tree-level Hagedorn temperature $\tilde\beta_\textrm{H}^{(0)}$
is obtained from the free partition function $Z^{(0)}$ which can be
written as \cite{Spradlin:2004pp,Harmark:2006di}
\begin{align}
\label{eq:Z0letter} \log Z^{(0)}(\tilde \beta,\tilde h) =
-\sum_{k=1}^{\infty}
    \log\left(1-z(k\tilde\beta, \tilde h) \right).
\end{align}
The letter partition function $z$ with general chemical potentials
was derived in \cite{Yamada:2006rx,Harmark:2006di} and in the
decoupling limit \eqref{declim} it reduces to
\begin{align}
\label{lettpart} z(\tilde \beta, \tilde h)
    = 2e^{-\tilde\beta} \cosh(\tilde\beta\tilde h).
\end{align}
The pole is located where the letter partition function goes to one,
as can be seen from equation \eqref{eq:Z0letter}. Note that it is
the $k=1$ pole that we are interested in since this is the first
pole that one encounters when raising the temperature. From
Eq.~\eqref{lettpart} we then find that the tree-level Hagedorn
inverse temperature $\tilde\beta^{(0)}_\textrm{H}$ is given by
Eq.~\eqref{zerohag}, as found from the Heisenberg chain through
Eq.~\eqref{genhag}.

The one-loop correction to the inverse Hagedorn temperature is given
by the residue at $\tilde\beta=\tilde \beta^{(0)}_\textrm{H}$ of the
first order contribution to the single-trace partition function
$\tilde\lambda Z^{(1)}_\textrm{ST}(\tilde \beta, \tilde h)$. This
first-order contribution is known to be given by
\cite{Spradlin:2004pp, Harmark:2006di}
\begin{align}
\label{eq:Z1ST} Z^{(1)}_\textrm{ST}(\tilde \beta, \tilde h)
  = -\tilde \beta \sum_{L=1}^\infty
    \frac{\langle D_2 (L \tilde \beta, \tilde h)\rangle}{1-z(L \tilde \beta, \tilde h)}
    + \textrm{(non-divergent $\langle PD_2 \rangle$ terms)}.
\end{align}
The expectation value $\langle D_2(\tilde \beta, \tilde h)\rangle$
is a special case of the more general expectation value that was
calculated in \cite{Harmark:2006di}. In our decoupling limit, it
does not depend on $\tilde h$ and is simply given by
\begin{align}
\langle D_2(\tilde \beta, \tilde h)\rangle = e^{-2\tilde\beta}.
\end{align}
The shift of the inverse Hagedorn temperature is therefore
\begin{align}
\delta \tilde \beta_{\rm {H}} = {\rm Res}_{\tilde \beta \to
\tilde\beta^{(0)}_{\rm {H}}} \left(\frac{ -\tilde\beta\tilde \lambda
e^{-2\tilde\beta}}{1-z(\tilde \beta, \tilde h)}\right)
\end{align}
which precisely gives the one-loop contribution \eqref{onehag} to
the Hagedorn temperature.

Thus, the more cumbersome method of computing the Hagedorn
temperature to order $\tilde{\lambda}$ explicitly from the pole of
the partition function gives the same result as computing it using
Eq.~\eqref{genhag}. This illustrates how powerful the relation
\eqref{genhag} is, also since it would be very hard to obtain higher
order corrections in $\tilde{\lambda}$ directly from the pole of the
partition function.

%%%%%%%%%%%%%%%%%%%%%%%%%%%%%%%%%%%%%%%%%%%%%%%%%%%%%%%%%%%%%%%%%%%%%%%%%%%%%%%%%%%%%%%%%%%%%%

\subsection{Spectrum with magnetic field for large $\tilde\lambda$ and large $L$}
\label{sec:lowspectrum}

In this section we consider the spectrum of the single-trace
operators of planar $\CN=4$ SYM on $\R \times S^3$ in the
decoupling limit \eqref{declim} for large $\tilde{\lambda}$ and
large $L$, $L$ being the length of the single-trace operators.

From Eqs.~\eqref{eq:logZ} and \eqref{zxxx} it is clear that the
effective Hamiltonian for the single-trace operators is $H = L +
H_{XXX}$, with $H_{XXX} = \tilde{\lambda} D_2 - 2 \tilde{h} S_z$. As
explained above, $H_{XXX}$ is the Hamiltonian for the Heisenberg
chain with coupling $\tilde{\lambda}$ and with an external magnetic
field of magnitude $2\tilde{h}$. We see therefore that the large
$\tilde{\lambda}$ regime corresponds to the low temperature regime
$\tilde{\beta} \tilde{\lambda} \gg 1$ of the Heisenberg chain. We
can therefore think of the spectrum for large $\tilde{\lambda}$ as
the low energy spectrum for the Heisenberg chain.

We explain now first how to find the large $\tilde{\lambda}$ and
large $L$ spectrum of single-trace operators by obtaining the low
energy spectrum of the Heisenberg chain in a non-zero magnetic field
$\tilde{h}$. After doing so, we discuss the physical difference from
the spectrum for $\tilde{h}=0$ and we explain why this difference
has important physical implications.

For $\tilde{h} > 0$ the vacuum is $\tr (Z^L)$ and we get the excited
states above the vacuum by inserting impurities in the form of $X$'s
into the $Z$'s in the vacuum. Considering the case of $M$
impurities, each with momentum $p_i$, $i=1,...,M$, the spectrum of
$H_{XXX}$ can be obtained using the Bethe ansatz technique together
with the integrability of the Heisenberg chain \cite{takahashi}. The
dispersion relation becomes
\begin{align}
\label{disprel} E= 2\tilde\lambda \sum_{i=1}^{M}
\sin^2\left(\frac{p_i}{2}\right)
    -\tilde h L + 2\tilde{h} M
\end{align}
where $E$ is the eigenvalue of $H_{XXX}$. The two terms with
$\tilde{h}$ arise from the $-2\tilde{h}S_z$ term in $H_{XXX}$. The
$M$ momenta $p_i$ are determined from the algebraic Bethe equations
\begin{align}
\label{betheeqs} e^{ip_kL} = \prod_{j=1,j\ne k}^{M} S(p_k,p_j),
\quad S(p_k,p_j) = \frac{1+e^{i(p_k+p_j)} -
2e^{ip_k}}{1+e^{i(p_k+p_j)} - 2e^{ip_j}}
\end{align}
together with the following condition coming from the cyclicity of
the trace
\begin{align}
\label{cyclcon} \sum_{i=1}^{M}p_i = 0.
\end{align}
At this point we did not make any approximation. However, we now
impose that we want the low energy spectrum in the thermodynamic
limit $L \rightarrow \infty$. This spectrum consists of the magnon
states, where each magnon corresponds to an impurity. In the low
energy approximation, the momenta $p_i$ of the magnons are taken to
be small. Also, we assume that $M \ll L$. From this we see that the
Bethe equations \eqref{betheeqs} to leading order become $e^{ip_k L}
= 1$, $k=1,...,M$. The leading order solution for the momenta is
therefore
\begin{align}
p_k = \frac{2\pi n_k}{L} + \mathcal{O}(L^{-2})
\end{align}
where $n_k$ is an integer. Inserting this in the dispersion relation
\eqref{disprel}, we get the spectrum of the magnons
\begin{equation}
E = -\tilde{h}L + \frac{2\pi^2\tilde \lambda}{L^2} \sum_{k=1}^M
n_k^2+2\tilde{h}M , \quad \sum_{k=1}^M n_k = 0,
\end{equation}
where the second equation is the cyclicity constraint
\eqref{cyclcon}. Defining now $M_n$ as the number of
impurities/magnons at momentum level $n$, i.e.\ how many of the
$n_k$'s are equal to $n$, we can write the spectrum as
\begin{equation}
E = -\tilde{h}L + \frac{2\pi^2\tilde \lambda}{L^2} \sum_{n\in
\mathbb{Z}} n^2M_n+2\tilde{h}\sum_{n \in \mathbb{Z}}M_n , \quad
\sum_{n\in \mathbb{Z}} n M_n = 0, \label{lowspec}
\end{equation}
where we used that the total number of impurities is $M=\sum_{n\in
\mathbb{Z}}M_n$.

Note that in the spectrum \eqref{lowspec} we have in particular the
mode $M_0$ that counts the number of impurities with zero momentum.
If we consider the states that have only zero-momentum impurities,
i.e.\ $M=M_0$, it is easily seen that they correspond to the totally
symmetrized single-trace operators
\begin{align}
\label{zeromodes} \tr(\mathrm{sym}(Z^{L-M}X^{M})).
\end{align}
Such operators are chiral primaries of $\CN=4$ SYM, so we have that
the vacuum $\tr(Z^L)$ and the zero modes \eqref{zeromodes} all
correspond to chiral primaries.

\subsubsection*{Breaking of the degeneracy of the vacuum by the
magnetic field}

It is well known that there is a significant difference between the
low energy spectrum of the ferromagnetic Heisenberg chain with or
without the external magnetic field.  With the magnetic field
present, the spins prefer to be aligned in the direction of the
field as the temperature goes to zero, but without $\tilde h$ there
is no preferred direction and the vacuum is degenerate. We explain
in the following how this manifests itself in our case.

We first review how the large $\tilde{\lambda}$ and large $L$
spectrum is found in the case of zero magnetic field $\tilde{h}=0$.
This case, corresponding to planar $\CN=4$ SYM on $\R \times S^3$ in
the decoupling limit \eqref{gaugelim}, is treated in Ref.\
\cite{Harmark:2006ta}. There is a degeneracy of the vacuum into the
$L+1$ different vacua
\begin{align}
\label{degvacua} \vert S_z\rangle_L \sim
\tr(\mathrm{sym}(Z^{\frac{1}{2}L + S_z}X^{\frac{1}{2}L - S_z}))
\end{align}
with each vacuum labeled by $S_z$ since the vacuum is degenerate
with respect to the total spin.

The low energy excitations above these vacua are magnons that can be
constructed using a novel approach to the Bethe ansatz technique
\cite{Harmark:2006ta}. Starting from each vacuum $\vert
S_z\rangle_L$, the magnons are made from impurities that correspond
to acting with the operator $S_z$ on particular sites of the chain.
In this way the total spin of the state does not change. The virtue
of this method is that low energy excitations can be studied above
any vacuum without running into finite size effects
\cite{Harmark:2006ta}.

With the external magnetic field present, however, the situation has
changed. The state $\vert S_z\rangle_L$ carries energy $-2\tilde h
S_z$ and therefore $\tr(Z^L)$, which has $S_z = L/2$, becomes the
unique vacuum and hence the $L+1$ fold degeneracy is removed. The
method of \cite{Harmark:2006ta} described above to build the low
energy states on top of the degenerate vacuum will therefore no
longer produce the correct spectrum. This can for example be seen by
considering a state $\vert S_z\rangle_L$ with $S_z \leq 0$. The
energy of such a state would at least be an energy $\tilde{h} L$
above the vacuum, which is outside the low energy regime that we are
considering (since $L$ is large). Thus, we cannot simply perturb
around the states obtained for $\tilde{h}=0$ in
\cite{Harmark:2006ta} to get the spectrum \eqref{lowspec} for
$\tilde{h}
> 0$. Therefore, the presence of the external magnetic field
$\tilde{h}$ has a non-trivial physical effect, even if it is close
to zero.

With the construction of the states where we insert $X$ as an
impurity%
\footnote{Note that inserting an impurity $X$ into $\tr(Z^L)$ at a
particular site can be seen as acting with the $SU(2)$ operator
$S_-$ at that site.} into the unique vacuum $\tr (Z^L)$, we find
instead the low energy spectrum without running into problems with
finite-size effects. We see also that the zero modes
\eqref{zeromodes} in this case correspond to the broken vacuum
states \eqref{degvacua} for the $\tilde{h}=0$ case.

%%%%%%%%%%%%%%%%%%%%%%%%%%%%%%%%%%%%%%%%%%%%%%%%%%%%%%%%%%%%%%%%%%%%%%%%%%%%%%%%%%%%%%%%%%%%%%
\subsection{Hagedorn temperature for large $\tilde \lambda$}
\label{sec:lowhag}

In this section we find the Hagedorn temperature $\tilde{T}_H$ of
planar $\CN=4$ SYM on $\R\times S^3$ in the decoupling limit
\eqref{declim} for large $\tilde{\lambda}$. The resulting
temperature $\tilde{T}_H$ depends on both $\tilde{\lambda}$ and
$\tilde{h}$. We compute $\tilde{T}_H$ by finding $V(\tilde{\beta})$
using the large $\tilde{\lambda}$ and large $L$ spectrum
\eqref{lowspec} derived in Section \ref{sec:lowspectrum}. The result
that we get for the Hagedorn temperature will be matched to the
Hagedorn temperature computed in string theory in Section
\ref{sec:string}.

From the spectrum \eqref{lowspec} we get that the partition function
for the Heisenberg chain for large $\tilde{\lambda}$ and large $L$
is given by
\begin{equation}
{\tr}_L \left( e^{- \tilde \beta H_{XXX} } \right) =
\sum_{\{M_n\}}\int_{-1/2}^{1/2}du \,e^{-\tilde \beta \left(
\frac{2\pi^2 \tilde \lambda}{ L^2} \sum_{n\in \mathbb{Z}} n^2
M_n+2\tilde{h}\sum_{n\in \mathbb{Z}}M_n-\tilde{h}L\right)+ 2\pi i u
\sum_{n\in \mathbb{Z}} n M_n }
\end{equation}
where the range in the sum over $M_n$ is from zero to infinity and
the cyclicity constraint in the spectrum \eqref{lowspec} is imposed
by introducing an integration over the variable $u$. After
evaluating the sums over the $M_n$'s we have
\begin{equation}
\label{Zsmallt} {\tr}_L \left( e^{- \tilde \beta H_{XXX}  }
\right) = \int_{-1/2}^{1/2} du \,e^{\tilde \beta
\tilde{h}L}\prod_{n=-\infty}^{\infty} \left[ 1- \exp \left( -
\frac{2\pi^2\tilde \beta \tilde \lambda}{L^2} n^2 -2\tilde \beta
\tilde{h}+ 2\pi i u n \right) \right]^{-1}.
\end{equation}
In order to obtain $V(\tilde{\beta})$, we should extract from
\eqref{Zsmallt} the part that diverges like $\exp (\mbox{const.}
\times L)$ for $L \rightarrow \infty$. It is possible to show that
the leading divergent contribution comes from $u=0$ and
that it is given by%
\footnote{For a detailed evaluation of the asymptotic behavior of
equation \eqref{Zsmallt} see \cite{Harmark:2006ta}. Note that in the
present case, contrary to the situation in \cite{Harmark:2006ta},
the product over $n$ extends from $-\infty$ to $\infty$ due to the
presence of the term proportional to $M_0$.}
\begin{equation}
\label{divcon} \exp\left[\tilde \beta
\tilde{h}L+L\sqrt{\frac{1}{2\pi\tilde \beta \tilde \lambda}}{\rm
Li}_{\frac{3}{2}}\left(e^{-2\tilde \beta\tilde{h}}\right)\right] \ \
\mbox{for} \ \ L \rightarrow \infty
\end{equation}
where ${\rm Li}_n(x)$ is the Polylogarithm function.%
\footnote{See for example App. E of \cite{Harmark:1999xt}.} Using
this result we can determine the expression for the function
$V(\tilde \beta)$ in the large $\tilde \lambda$ limit which reads
\begin{equation}
\label{Vlowt} V(\tilde \beta) =\tilde \beta
\tilde{h}+\sqrt{\frac{1}{2\pi\tilde \beta \tilde \lambda}}{\rm
Li}_{\frac{3}{2}}\left(e^{-2\tilde \beta\tilde{h}}\right).\ \
\end{equation}
This gives the thermodynamics of the Heisenberg chain with
Hamiltonian \eqref{ham} in the low temperature
$\tilde{\beta}\tilde{\lambda} \ll 1$ and large $L$ limit. To the
best of our knowledge, this is a new result for the magnetic
Heisenberg chain.

Inserting now the result \eqref{Vlowt} in \eqref{genhag} we get the
following equation for the Hagedorn temperature
\begin{equation}
\tilde T_H=\left[(1-\tilde h)\sqrt{2\pi\tilde \lambda}\left({\rm
Li}_{\frac{3}{2}}(e^{-2\tilde{h}/\tilde
T_H})\right)^{-1}\right]^{2/3} \label{Hagh}.
\end{equation}
Note that for $\tilde h=0$ we recover the result for the large
$\tilde{\lambda}$ Hagedorn temperature recently obtained in
\cite{Harmark:2006ta}.

\begin{figure}[t]
\centerline{\epsfig{file=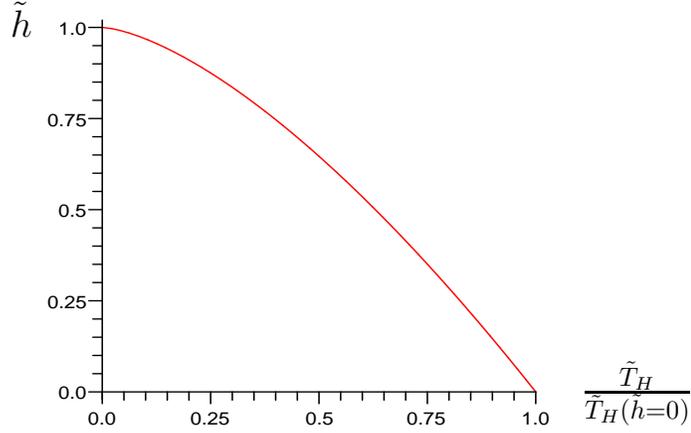,width=8cm,height=7cm} }
\caption{The Hagedorn temperature $\tilde{T}_H$ as function of
$\tilde{h}$ for large $\tilde{\lambda}$. Here $\tilde{T}_H
({\tilde{h}=0}) = (2\pi)^{1/3} (\zeta(3/2))^{-2/3}
\tilde{\lambda}^{1/3}$. \label{fig:htcurve_strong}}
\begin{picture}(0,0)(0,0)
\put(117,233){\Large $\tilde h$} \put(333,95){\Large
$\frac{\tilde{T}_H}{\tilde{T}_H ({\tilde{h}=0})}$}
\end{picture}
\end{figure}

From Eq.~\eqref{Hagh} we see that the Hagedorn temperature goes like
$\tilde \lambda^{1/3}$ to leading order so it is large for large
$\tilde \lambda$. This means that $\tilde{h}/\tilde{T}_H$ is small
since $0 \leq \tilde{h} \leq 1$ and it makes sense therefore to
expand the Polylogarithm function. This gives the following result
for $\tilde{T}_H$ as a function of $\tilde{h}$ for large
$\tilde{\lambda}$
\begin{equation}
\tilde{T}_H = \frac{ (2\pi)^{\frac{1}{3}} ( 1-\tilde
h)^{\frac{2}{3}}} { \zeta(\frac{3}{2})^{\frac{2}{3}} }
\tilde{\lambda}^{\frac{1}{3}}+\frac{4(2\pi)^{\frac{2}{3}}\sqrt{\tilde
h}(1-\tilde h)^{\frac{1}{3}}}{3\, \zeta(\frac{3}{2})^{\frac{4}{3}}}
\,\tilde{\lambda}^{\frac{1}{6}}+\mathcal{O}(\tilde\lambda^0).\label{Hagh12}
\end{equation}
Note that at order $\tilde{\lambda}^0$ there are other corrections
to the spectrum \eqref{lowspec} that must be taken into account
\cite{Harmark:2006ta}.

An interesting feature of \eqref{Hagh12} is that $\tilde{T}_H
\rightarrow 0$ as $\tilde{h} \rightarrow 1$. This is the same as in
the case of small $\tilde{\lambda}$, as found in Section
\ref{sec:smalllamb}. In Fig.~\ref{fig:htcurve_strong} we have
plotted the leading behavior of $\tilde{T}_H$ as a function of
$\tilde{h}$ for large $\tilde{\lambda}$, i.e.\ the first term in
\eqref{Hagh12}. It is interesting to compare this curve to the one
in Fig.~\ref{fig:htcurve_weak} for $\tilde{\lambda}=0$. We expect
that the shape of the curve will interpolate smoothly between the
small and large $\tilde{\lambda}$ regimes.

%%%%%%%%%%%%%%%%%%%%%%%%%%%%%%%%%%%%%%%%%%%%%%%%%%%%%%%%%%%%%%
\section{String theory side: The magnetic pp-wave}
\label{sec:string}

In this section we consider the dual string theory version of the
decoupling limit \eqref{declim} of $\CN=4$ SYM on $\R \times S^3$.
We write down the limit in the microcanonical ensemble, which is
appropriate for taking the limit on the string side. We then go on
to find a Penrose limit which can give a pp-wave background
compatible with the decoupling limit. After implementing the
decoupling limit for the pp-wave, we match the spectrum and the
Hagedorn temperature of weakly coupled strings to the corresponding
quantities on the gauge theory side as obtained in Section
\ref{sec:spinchain}.

%%%%%%%%%%%%%%%%%%%%%%%%%%%%%%%%%%%%
\subsection{Decoupling limit of strings on $\ads_5 \times S^5$}
\label{sec:decads}

In order to find the dual decoupling limit for strings on
$\ads_5\times S^5$ we should first reformulate the decoupling limit
\eqref{declim} of thermal $\CN = 4$ SYM on $\R \times S^3$ as a
decoupling limit that does not refer to temperature, i.e.\ a
decoupling limit in the microcanonical ensemble. This can be done by
analyzing the weight factor in the partition function
\eqref{eq:grandcan} which we can write as
\begin{equation}
e^{-\beta(1-\Omega)J  -\beta (D-J) + 2 \beta h S_z}.
\end{equation}
With $h=0$ we can implement the decoupling limit \eqref{declim} in
the microcanonical ensemble by considering $\tilde{H} =
(D-J)/\epsilon$ fixed and $\lambda /\epsilon$ fixed, and then taking
$\epsilon \rightarrow 0$ \cite{Harmark:2006ta}. However, we see here
that the presence of the extra term means that we instead should
rescale $D-J - 2hS_z +h J = D-J_1 -(1-2h) J_2$, where we have added
the term $hJ$ so that the vacuum has energy zero. It is important to
remember that $h$ is also rescaled (see \eqref{declim}), which is
necessary to have the decoupling of the states in the $SU(2)$
sector. The decoupling limit of $\CN = 4$ SYM on $\R \times S^3$ in
the microcanonical ensemble can thus be written as
\begin{equation}
\label{microdeclim} \epsilon \rightarrow 0,\quad \tilde{H} \equiv
\frac{D-J_1-(1-2h)J_2}{\epsilon} \ \mbox{fixed}, \quad
\tilde{\lambda} \equiv \frac{\lambda}{\epsilon} \ \mbox{fixed} ,
\quad \tilde{h} \equiv \frac{h}{\epsilon} \ \mbox{fixed} , \quad
J_i, N \ \mbox{fixed}.
\end{equation}
This is the limit that we should translate to a decoupling limit of
the dual string theory. Note that we have in the decoupling limit
that
\begin{equation}
\label{Htilde} \tilde{H} = \tilde{h} J  + \tilde{\lambda} D_2 -
2\tilde{h} S_z = \tilde{h} J  +  H_{XXX}
\end{equation}
with $H_{XXX}$ defined as in \eqref{ham}.

On the string theory side, we are considering type IIB string theory
on the $\ads_5 \times S^5$ background given by the metric
\begin{equation}
\label{adsmet} ds^2 = R^2 \left[ - \cosh^2 \rho dt^2 + d\rho^2 +
\sinh^2 \rho d{\Omega'_3}^2 + d\theta^2 + \sin^2 \theta d\alpha^2 +
\cos^2 \theta d\Omega_3^2 \right]
\end{equation}
and the five-form Ramond-Ramond field strength
\begin{equation}
\label{adsF5} F_{(5)} = 2 R^4 ( \cosh \rho \sinh^3 \rho dt d\rho
d\Omega_3' +
 \sin \theta \cos^3 \theta d\theta d\alpha d\Omega_3 ).
\end{equation}
Here the radius $R$ is given by $R^4 = 4\pi g_s l_s^4 N$ and $\gym^2
= 4\pi g_s$, where $g_s$ is the string coupling and $l_s$ is the
string length. Note that $\gym$ and $N$ are the gauge coupling and
rank of $SU(N)$ as defined in Section \ref{sec:declim}. With this,
we see that we have the following dictionary between the gauge
theory quantities $\lambda$ and $N$, and the string theory
quantities $g_s$, $l_s$ and the AdS radius $R$
\begin{equation}
\label{adsdic} T_{\rm str} \equiv \frac{R^2}{4\pi l_s^2} =
\frac{1}{2} \sqrt{\lambda}, \quad g_s = \frac{\pi \lambda}{N}
\end{equation}
where $T_{\rm str}$ is the string tension for a fundamental string
in the $\ads_5 \times S^5$ background \eqref{adsmet}--\eqref{adsF5}.

In the following we write $E$ for the energy of the string. The
energy $E$ for a string state is dual to the scaling dimension $D$
of a gauge theory state of $\CN=4$ SYM on $\R \times S^3$ since we
set the radius of the three-sphere to one.  We furthermore write
$J_i$, $i=1,2,3$, for the three angular momenta on the five-sphere
dual to the three R-charges of $\CN=4$ SYM, and we write $\Omega_i$,
$i=1,2,3$, as the corresponding angular velocities, dual to the
chemical potentials for the R-charges of $\CN=4$ SYM.

We can now translate the decoupling limit \eqref{microdeclim} into
the following limit of type IIB string theory on the $\ads_5 \times
S^5$ background \eqref{adsmet}--\eqref{adsF5}
\begin{equation}
\label{microadslim}
\begin{array}{c} \ds
\epsilon \rightarrow 0 ,\quad
\tilde{H} \equiv \frac{E-J_1-(1-2h)J_2}{\epsilon} \ \mbox{fixed}, \quad
\tilde{T}_{\rm str} \equiv \frac{T_{\rm str}}{\sqrt{\epsilon}} \ \mbox{fixed},
\\[3mm]
\ds \tilde{g}_s \equiv \frac{g_s}{\epsilon}\ \mbox{fixed}, \quad
\tilde{h} \equiv \frac{h}{\epsilon} \ \mbox{fixed}, \quad
J_i \ \mbox{fixed}.
\end{array}
\end{equation}
This limit closely resembles the decoupling limit of
strings on $\ads_5\times S^5$ found in \cite{Harmark:2006ta}. The
only difference is the deformation caused by the $h$ parameter. This
adds an extra term to the effective Hamiltonian $\tilde{H}$ for the
strings. We also see that we get the following dictionary between
the gauge theory and string theory quantities in the respective
decoupling limits \eqref{microdeclim} and \eqref{microadslim}
\cite{Harmark:2006ta}
\begin{equation}
\label{redadsdic}
\tilde{T}_{\rm str} = \frac{1}{2}\sqrt{\tilde{\lambda}} , \quad
\tilde{g}_s = \frac{\pi\tilde{\lambda}}{N}
\end{equation}
which mirrors the AdS/CFT dictionary \eqref{adsdic}.

In order to fully justify that $\tilde{H}$ in \eqref{microadslim} is
the right expression for the effective Hamiltonian we should
consider a thermal gas of strings in the $\ads_5 \times S^5$
background \eqref{adsmet}--\eqref{adsF5}. We can write the general
partition function as
\begin{equation}
Z(\beta,\Omega_i) = \tr \left( e^{-\beta E + \beta \sum_{i=1}^3
\Omega_i J_i} \right).
\end{equation}
Putting $\Omega_3=0$ and writing $\Omega_1 = \Omega+h$, $\Omega_2 =
\Omega-h$, $J=J_1+J_2$ and $S_z = (J_1-J_2)/2$, as on the gauge
theory side, we get
\begin{equation}
\label{partff} Z(\beta,\Omega,h) = \tr \left( e^{-\beta E + \beta
\Omega J + 2 \beta h S_z} \right) = \tr \left( e^{-\beta [
E-J_1-(1-2h)J_2 ] - \beta (1-\Omega-h) J} \right).
\end{equation}
Taking now the limit \eqref{microadslim} with
\begin{equation}
\label{epsom} \epsilon = 1-\Omega , \quad \tilde{T} \equiv
\frac{T}{1-\Omega},
\end{equation}
we see that the partition function \eqref{partff} reduces to
\begin{equation}
\label{zzz} Z(\tilde{\beta},\tilde{h}) = \tr \left\{ \exp \left[
-\tilde{\beta} \left( (1-\tilde{h}) J + \tilde{H} \right) \right]
\right\}
\end{equation}
where the trace is now over a reduced set of string theory states,
corresponding to the decoupling of the $SU(2)$ sector on the gauge
theory side. We see from this that the total Hamiltonian is
$(1-\tilde{h}) J + \tilde{H}$, thus for a fixed $J$ we can regard
$\tilde{H}$ as the effective Hamiltonian. We note finally that
$\tilde{h} \leq 1$ since otherwise the partition function
\eqref{zzz} is not well-defined.

%%%%%%%%%%%%%%%%%%%%%%%%%%%%%%%%%%%%%%%%%%
\subsection{Finding the Penrose Limit}
\label{sec:findpenrose}

The goal of this section is to find a Penrose limit of $\ads_5
\times S^5$ which gives a pp-wave background matching the large
$\tilde{\lambda}$ spectrum and Hagedorn temperature found on the
gauge theory side.

In the following we parameterize the three-sphere inside the
five-sphere in \eqref{adsmet} as
\begin{equation}
\label{3sph} d\Omega_3^2 = d\psi^2  + \cos^2 \psi d\chi^2 + \sin^2
\psi d\phi^2 .
\end{equation}
We define that $J_1 = - i \partial_\chi$, $J_2 = -i
\partial_\phi$ and $J_3 = - i \partial_\alpha$.
{}From Section \ref{sec:decads} we have that the effective
Hamiltonian for which the vacuum state has zero energy is
proportional to $E - J_1 - (1-2h)J_2$. In accordance with previously
found Penrose limits (see Appendix \ref{app:penrose} for the BMN
Penrose limit \cite{Berenstein:2002jq} and the Flat Direction
Penrose limit \cite{Bertolini:2002nr}) this means that we should
consider two new variables $\phi_+$ and $\phi_-$ defined in terms of
the five-sphere coordinates $\chi$ and $\phi$ such that $J_+ \equiv
- i
\partial_{\phi_+} = J_1 + (1-2h)J_2$ since this gives that the
Hamiltonian is proportional to $E - J_+$. Thus, we should require
\begin{equation}
\label{currcond}
\partial_{\phi_+} = \partial_\chi + (1-2h) \partial_\phi \,.
\end{equation}
The most general linear relation between $\phi_+$, $\phi_-$ and
$\chi$, $\phi$ obeying \eqref{currcond} is
\begin{equation}
\chi = \phi_+ + c_1 \phi_- \; , \quad \phi = (1-2h) \phi_+ + c_2 \phi_-\; .
\end{equation}
We see from Appendix \ref{app:penrose} that the BMN Penrose limit
\cite{Berenstein:2002jq} corresponds to $h=1/2$, $c_1 = 0$ and
$c_2=1$, while the Flat Direction Penrose limit
\cite{Bertolini:2002nr} corresponds to $h=0$, $c_1=1$ and $c_2=-1$.

Define now, as in \cite{Harmark:2006ta}, the rescaled AdS radius
$\tilde{R}$ by
\begin{equation}
\label{Rtilde} \tilde{R}^4 = \frac{R^4}{\epsilon}.
\end{equation}
The rescaled radius $\tilde{R}$ is fixed in the decoupling limit
\eqref{microadslim}. The light-cone coordinates are defined as
\begin{equation}
\label{zplmi} z^+ = \frac{1}{2\mu} ( t + \phi_+ ) , \quad
z^- = \frac{1}{2} \mu \tilde{R}^2 ( t - \phi_+ ),
\end{equation}
where the mass parameter $\mu$ has been introduced for later
convenience.  The Penrose limit will then consist of taking the
$\tilde{R}\rightarrow \infty$ limit. We now want to examine which
choices of $c_1$ and $c_2$ can lead to a consistent Penrose
limit.

Consider the following part of the the $\ads_5\times S^5$ metric
\eqref{adsmet}
\begin{equation}
\label{met1} \tilde{R}^2 ( -dt^2 + \sin^2 \psi d\phi^2 + \cos^2 \psi
d\chi^2 ) .
\end{equation}
This is the only part of the metric where we can get $dz^+$ terms.
Note that here and in the following we ignore the $\sqrt{\epsilon}$
factor in front of the metric since it will not be of importance for
these considerations. Considering now only $(dz^+)^2$ terms in
\eqref{met1}, we get
\begin{equation}
- \mu^2 \tilde{R}^2 \left[ 1 - (1-2h) \sin^2 \psi - \cos^2 \psi
\right] (dz^+)^2 .
\end{equation}
Since this is of order $\tilde{R}^2$, we need that $(1-2h) \sin^2
\psi + \cos^2 \psi =1$ to leading order in $1/\tilde{R}$, in order
to have a well-defined Penrose limit. However, this is equivalent to
demanding that
\begin{equation}
h \sin^2 \psi = 0.
\end{equation}
This is possible only if either $h=0$ or $\sin \psi = 0$. Thus, if
we want a background with $h > 0$ we are bound to impose that $\sin
\psi =0$ to leading order in $1/\tilde{R}$. On the other hand, with
$h=0$ we can freely choose $\psi$, and for the Flat Direction
Penrose limit (see Appendix \ref{app:penrose}) this is used to
choose $\psi = \pi/4$ to leading order. Therefore, any Penrose limit
giving a background with $h > 0$ will necessarily be disconnected
from the Flat Direction limit no matter how small $h$ is. This can
be understood as a geometric realization of the symmetry breaking
caused by the magnetic field in the ferromagnetic Heisenberg spin
chain as discussed in Section \ref{sec:lowspectrum}. For the spin
chain it is well known that an arbitrarily small magnetic field can
change the vacuum structure of the spin chain, and thereby also the
low energy spectrum. In the context of Penrose limits of
AdS$_5\times S^5$, however, this is a new result.

Thus, we can conclude from the above that since we want a Penrose
limit with $h > 0$ we should have $\psi \rightarrow 0$ in the
Penrose limit. Consider therefore the following part of the
$\ads_5\times S^5$ metric \eqref{adsmet} for $\rho=\theta=\psi=0$
\begin{equation}
\tilde{R}^2 ( -dt^2 + d\chi^2 ).
\end{equation}
In terms of $z^\pm$ and $\phi_-$, this metric is
\begin{equation}
- 4 dz^+ dz^- + c_1^2 \tilde{R}^2 d\phi_-^2 + 2 c_1 \mu
\tilde{R}^2 dz^+ d\phi_- - \frac{2c_1}{\mu} dz^- d\phi_- \, .
\end{equation}
If $c_1 \neq 0$, then the third terms means that we should have
$\phi_-$ of order $1/\tilde{R}^2$ or of higher order.%
\footnote{Note that we assume $c_1$ to be independent of
$\tilde{R}$. One could also imagine having $c_1$ of order
$1/\tilde{R}^2$. This, however, does not lead to any interesting new
limits.} However, this has the consequence that $d\phi_-$ does not
appear in any other part of the metric after the Penrose limit,
which clearly is not consistent. Therefore, we can conclude that we
should restrict ourselves to having $c_1=0$.

Now that $c_1=0$, we can choose our normalization for $\phi_-$ such
that $c_2=1$. This is a useful choice since it means that $J_-
\equiv - i
\partial_{\phi_-} = J_2$. From our considerations we can therefore
fix that
\begin{equation}
\label{chiphi} \chi = \phi_+ \; , \quad \phi = (1-2h) \phi_+ +  \phi_- \; .
\end{equation}
We can now write down the Penrose limit. Defining
\begin{equation}
\label{rdefs} r = \tilde{R} \rho , \quad \tilde{r} = \tilde{R}
\theta , \quad \bar{r} = \tilde{R} \psi ,
\end{equation}
the Penrose limit is
\begin{equation}
\label{penlim} \tilde{R} \rightarrow \infty, \quad
z^+, z^-, r, \Omega_3',\tilde{r},\alpha,\bar{r},\phi_- \ \mbox{fixed},
\end{equation}
where the coordinates listed are defined in \eqref{adsmet},
\eqref{zplmi}, \eqref{chiphi} and \eqref{rdefs}. The Penrose limit
\eqref{penlim} of the $\ads_5\times S^5$ background
\eqref{adsmet}--\eqref{adsF5} results in the following pp-wave
background with metric
\begin{equation}
\label{ppwmet}
\begin{array}{rcl} \ds
\frac{1}{\sqrt{\epsilon}} ds^2 &=&\ds - 4 dz^+ dz^- - \mu^2 \left(
(1-(1-2h)^2) \sum_{i=1}^2 (z^i)^2 + \sum_{I=3}^8 (z^I)^2 \right)
(dz^+)^2 \\[4mm] && \ds + \sum_{i=1}^8 (dz^i)^2 + 2 \mu (1-2h)
\left[ z^1 dz^2 - z^2 dz^1 \right] dz^+
\end{array}
\end{equation}
and five-form field strength
\begin{equation}
\label{ppwF5} \frac{1}{\epsilon} F_{(5)} = 2\mu dz^+ (dz^1 dz^2 dz^3
dz^4 + dz^5 dz^6 dz^7 dz^8 ).
\end{equation}
Here $\mu$ is the mass parameter introduced in \eqref{zplmi}. The
coordinates $z^1,z^2$ are defined by $z^1+iz^2 = \bar{r}
e^{i\phi_-}$, $z^3,z^4$ are defined by $z^3+iz^4 = \tilde{r}
e^{i\alpha}$ and $z^5,...,z^8$ are defined by $r^2 = \sum_{i=5}^8
(z^i)^2$ and $dr^2+r^2 d\Omega_3^2 = \sum_{i=5}^8 (dz^i)^2$.

It is important to note that the Penrose limit \eqref{penlim}
becomes the BMN Penrose limit \eqref{BMNlimit}
\cite{Berenstein:2002jq} if we set $h=1/2$. Moreover, as a
consequence of this, the resulting pp-wave background
\eqref{ppwmet}--\eqref{ppwF5} is seen to reduce to
\eqref{BMNmet}--\eqref{BMNF5} for $h=1/2$.

Considering now the Penrose limit \eqref{penlim} in terms of the
generators, we have the relations
\begin{equation}
\label{thegen} H_{\rm lc} = \sqrt{\epsilon} \mu ( E - J_+ ) , \quad
p^+ = \frac{E+J_+}{2\mu R^2},
\end{equation}
where $H_{\rm lc}$ is the light-cone Hamiltonian and $p^+$ is the
light-cone momentum. It follows from these relations that the
Penrose limit \eqref{penlim} is such that $J_+/\tilde{R}^2$ and
$E-J_+$ are fixed in the limit $\tilde{R} \rightarrow \infty$. In
particular, this means that $J_+ = J_1 + (1-2h)J_2 \rightarrow
\infty$. However, since $J_- = J_2$ and since we keep $\phi_-$
fixed, we have that $J_2$ is fixed in the Penrose limit
\eqref{penlim}. Therefore, in terms of $E$, $J_1$ and $J_2$, the
Penrose limit \eqref{penlim} corresponds to taking the limit
\begin{equation}
\label{penlim2}
\tilde{R} \rightarrow \infty , \quad
J_1 \rightarrow \infty, \quad
\frac{J_1}{\tilde{R}^2} \ \mbox{fixed}, \quad
J_2\ \mbox{fixed}, \quad
 E-J_1 \ \mbox{fixed}.
\end{equation}
We see that this is the same limit of the generators as that
corresponding to the BMN Penrose limit \cite{Berenstein:2002jq}. We
are thus considering the same set of string states in the Penrose
limit \eqref{penlim} as in the BMN Penrose limit. This is contrary
to the Flat Direction Penrose limit \cite{Bertolini:2002nr} which
involves a different set of string states. We see therefore that
even though $h$ can be arbitrarily close to zero, the Penrose limit
concerns the same set of string states as for the BMN Penrose limit
with $h=1/2$, despite the fact that the Flat Direction Penrose limit
is the relevant one for $h=0$. This is another manifestation of the
symmetry breaking caused by the magnetic field discussed in Section
\ref{sec:lowspectrum}.

If we compare the background \eqref{ppwmet}--\eqref{ppwF5} to the
pp-wave background
\eqref{eq:transformedmetric}--\eqref{eq:transformedF5} found in
Appendix \ref{app:ppwave} by a constant rotation of the BMN pp-wave
background \eqref{BMNmet}--\eqref{BMNF5} we see that
\eqref{ppwmet}--\eqref{ppwF5} corresponds to the pp-wave background
\eqref{eq:transformedmetric} for
\begin{equation}
\label{ideta} \eta = 1- 2h , \quad C = 0 .
\end{equation}
See Appendix \ref{app:ppwave} for a first quantization of type IIB
string theory in the pp-wave background
\eqref{eq:transformedmetric}--\eqref{eq:transformedF5}, leading to
the string spectrum \eqref{fullspec1} with level-matching condition
\eqref{fullspec2}.

As we discuss in Appendix \ref{app:ppwave} there are two ways to
think about the background \eqref{ppwmet}--\eqref{ppwF5}. We can
either think of it as the BMN pp-wave background rotated with a
constant angular velocity in one of the two-planes. This makes sense
since $h = (\Omega_1-\Omega_2)/2$ and since $\Omega_i$ are angular
velocities. Taking the limit $h \rightarrow 0$ as in
\eqref{microadslim} then means that we are approaching the critical
angular velocity $\eta=1$.

Furthermore, as discussed in Section \ref{app:physint}, we can also
think about the pp-wave background \eqref{ppwmet}--\eqref{ppwF5} as a
magnetic pp-wave background, in the sense that the Hamiltonian for
the background is equivalent to that of a particle in a constant
magnetic field along with a harmonic oscillator potential. This is
interesting since we precisely are turning on a magnetic field for
the Heisenberg spin chain, and we can thus say that we have a
correspondence between the magnetic Heisenberg spin chain and the
magnetic pp-wave. However, the analogy is not perfect since the
pp-wave can also be said to be magnetic for $h=0$.

%%%%%%%%%%%%%%%%%%%%%%%%%%%%%%%%%%%%
\subsection{Decoupling limit of the pp-wave and matching of spectra}
\label{sec:decpp}

We now implement the decoupling limit \eqref{microadslim} for type
IIB strings on $\ads_5\times S^5$ on the pp-wave background
\eqref{ppwmet}--\eqref{ppwF5}. Since we want to keep $p^+$ as given
in \eqref{thegen} fixed in the decoupling limit we see that we need
$\sqrt{\epsilon}\mu$ to be kept fixed, like in
\cite{Harmark:2006ta}. Therefore, we get that the decoupling limit
for type IIB strings on the pp-wave background
\eqref{ppwmet}--\eqref{ppwF5} is given by
\begin{equation}
\label{pplim}
\epsilon \rightarrow 0 , \quad
\tilde{\mu} \equiv \mu \sqrt{\epsilon}\ \mbox{fixed}, \quad
\tilde{H}_{\rm lc} \equiv \frac{H_{\rm lc}}{\epsilon}\ \mbox{fixed}, \quad
\tilde{h} \equiv \frac{h}{\epsilon}\ \mbox{fixed} , \quad
\tilde{g}_s \equiv \frac{g_s}{\epsilon}\ \mbox{fixed} , \quad
l_s,\ p^+ \ \mbox{fixed}.
\end{equation}
We see that this limit reduces to the one of \cite{Harmark:2006ta}
for $\tilde{h}=0$. Clearly the limit \eqref{pplim} is a large $\mu$
limit of the magnetic pp-wave background
\eqref{ppwmet}--\eqref{ppwF5}. It is important to note here that we
have the bound $0 \leq \tilde{h} \leq 1$, where the upper bound is
discussed above, and the lower bound comes from the fact that $h$ is
required to be positive from the bound $|\eta|\leq 1$ and
\eqref{ideta}.

We obtain in Appendix \ref{sec:fullspec} the spectrum
\eqref{fullspec1}--\eqref{fullspec2} for the pp-wave background
\eqref{ppwmet}--\eqref{ppwF5}, with $\eta$ given by \eqref{ideta}.
Taking then the decoupling limit \eqref{pplim}, we get the reduced
spectrum
\begin{equation}
\label{spec1} \frac{1}{\tilde{\mu}} \tilde{H}_{\rm lc} = \frac{1}{2
(\tilde{\mu} l_s^2 p^+ )^2 } \sum_{n \neq 0} n^2 M_n +2 \tilde h
\sum_{n=-\infty}^{\infty} M_n, \quad \quad \sum_{n\neq 0} n M_n = 0,
\end{equation}
where we also included the level matching condition. This is the
spectrum for string theory on $\ads_5\times S^5$ in the decoupling
limit \eqref{microadslim} for large $\tilde{R}$ and large $J_1$.
Note furthermore that from \eqref{thegen} we have
\begin{equation}
\label{redpplus} \tilde{\mu} l_s^2 p^+ = \frac{J_1}{4\pi
\tilde{T}_{\rm str}}
\end{equation}
so we are in a region with large $\tilde{T}_{\rm str}$ and $J_1$. It
is interesting to observe from the spectrum \eqref{spec1} that the
string theory effectively becomes one-dimensional in the sense that
only the $M_n$ modes survive the limit \eqref{pplim}.

We now translate our results for the string theory side to gauge
theory, to examine the matching with the result for the gauge theory
side in Section \ref{sec:spinchain}. From Eq.~\eqref{redadsdic} we
see that the Penrose limit \eqref{penlim2} corresponds to the
following region of the decoupled gauge theory
\begin{equation}
\label{penlimreg}
\tilde{\lambda} \rightarrow \infty , \quad
J_1 \rightarrow \infty , \quad
\frac{\tilde{\lambda}}{J_1^2} \ \mbox{fixed}, \quad
J_2 \ \mbox{fixed}.
\end{equation}
Thus, we should match the spectrum \eqref{spec1} to the gauge theory
spectrum for large $\tilde{\lambda}$ and large $L=J_1+J_2$, which is
computed in Section \ref{sec:lowspectrum}. That $J_2$ is fixed in
the Penrose corresponds to the fact that we are inserting $X$ as
impurities in the ground state $\tr(Z^L)$ on the gauge theory side.
Therefore, since $J_2$ is the number of impurities it is consistent
with the gauge theory side that it is unaffected by the Penrose
limit. Note also that this is consistent with our low energy
approximation on the spin chain side in which we demand that the
number of impurities is not large, i.e.\ $J_2 \ll L$, since
otherwise one runs into finite size effects.

Now that we have established that the Penrose limit regime
\eqref{penlimreg} is in accordance with the low energy regime
considered in Section \ref{sec:lowspectrum}, we are left with
checking explicitly that the spectra \eqref{lowspec} and
\eqref{spec1} agree. To see this, we first note that one should
identify $\tilde{H}$ in \eqref{microdeclim} and \eqref{microadslim}.
Using then \eqref{Htilde} we see that we should make the
identification
\begin{equation}
\frac{1}{\tilde{\mu}} \tilde{H}_{\rm lc} = \tilde{h} L + H_{XXX}
\end{equation}
between the string theory and the gauge theory/spin chain energies.
Using then \eqref{redpplus} and \eqref{redadsdic} we see that the
spectrum \eqref{spec1} matches the spectrum \eqref{lowspec}
computed for planar $\CN=4$ SYM on $\R \times S^3$ in the decoupling
limit \eqref{declim} for large $\tilde{\lambda}$ and large $L$. Note
that as part of this matching we use that $J_1 \simeq L$ since $J_2
\ll L$.

%%%%%%%%%%%%%%%%%%%%%%%%%%%%%%%%%%%%%%%%%%%%%%%%%%%%%%%%%%%%%%%%%%%%%%%%%%%%%%%%%%%%%%%%%%%

\subsection{Hagedorn temperature on the string side}
\label{sec:hagstring}

In this section we compute the Hagedorn temperature for strings on
the pp-wave background \eqref{ppwmet}--\eqref{ppwF5} in the
decoupling limit \eqref{microadslim}, \eqref{pplim}. The
computation is done in two ways. First we compute the Hagedorn
temperature using the reduced pp-wave spectrum \eqref{spec1} and
subsequently we instead compute the Hagedorn temperature from the
full pp-wave spectrum \eqref{fullspec1}--\eqref{fullspec2} and
then we take the decoupling limit \eqref{pplim} on the result. We
show that in both cases we get the same result, which, moreover,
can be successfully matched with the Hagedorn temperature
\eqref{Hagh} computed on the gauge theory/spin-chain side. Note
that on the gauge theory side we have weakly coupled $\CN=4$ SYM.

The Hagedorn temperature of type IIB string theory on the maximally
supersymmetric pp-wave background of \cite{Blau:2001ne} has
previously been computed in
\cite{PandoZayas:2002hh,Greene:2002cd,Sugawara:2002rs,
Brower:2002zx,Sugawara:2003qc,Grignani:2003cs,Hyun:2003ks,Bigazzi:2003jk}.
We begin by
considering the multi-string partition function%
\footnote{For related computations of the string theory partition
function and Hagedorn temperature in the presence of background
fields, that play the role of chemical potentials for the
corresponding momenta, see for example Refs. \cite{Deo:1989bv,
Greene:2002cd}.}
\begin{equation}
\log Z(a,b,\mu, h)=\sum_{n=1}^{\infty}\frac{1}{n}\mbox{Tr}\left(
(-1)^{(n+1) {\rm \bf F}} e^{-a n H_{{\rm
lc}}-bnp^+}\right)\label{fab2}
\end{equation}
where the trace is over single-string states with the spectrum
\eqref{fullspec1}--\eqref{fullspec2}, and ${\rm \bf F}$ is the
space-time fermion number. The parameters $a$ and $b$ are introduced
as the inverse temperature and chemical potential for strings on the
pp-wave background \eqref{ppwmet}--\eqref{ppwF5}.

We have seen that in the decoupling limit \eqref{pplim} most of
the states decouple and the resulting pp-wave light-cone string
spectrum is given by eq.~\eqref{spec1}. We see that only the term
proportional to the bosonic modes $M_n$ contributes to the
spectrum in the limit \eqref{pplim}. We introduce therefore the
``reduced" multi-string partition function
\begin{equation}
\log Z (\tilde{a},\tilde{b},\tilde{\mu},\tilde h) =
\sum_{n=1}^{\infty} \frac{1}{n} \mbox{Tr} \left( e^{-\tilde{a} n
\tilde H_{{\rm lc}}-\tilde{b}np^+}\right) \label{fab}
\end{equation}
where the trace is now taken over single-string states with
spectrum \eqref{spec1} and $\tilde a$ and $\tilde b$ are the
inverse temperature and chemical potential after the limit
\eqref{pplim}. The computation of the Hagedorn temperature then
proceeds similarly to the one of Section \ref{sec:lowhag}. We
obtain that the Hagedorn singularity is defined by the equation
\begin{equation}
\tilde{b} \sqrt{\tilde{a}} =  l_s^2 \sqrt{2\pi\tilde{\mu}}\,{\rm
Li}_{3/2}\left(e^{-2\tilde a\tilde \mu \tilde h}\right)
\label{hageab}
\end{equation}
where ${\rm Li}_n(x)$ is the Polylogaritm function. We now
identify $\tilde{a}$ and $\tilde{b}$ in terms of the thermal
partition function \eqref{zzz} for a thermal gas of strings in the
$\ads_5\times S^5$ background in the decoupling limit
\eqref{microadslim}. This is done using \eqref{thegen} and
\eqref{pplim}, along with \eqref{redpplus} and the fact that
$J_1 \simeq J$. The result is
\begin{equation}
\label{ab} \tilde{a} = \frac{1}{\tilde{\mu}}\tilde{\beta} , \quad
\tilde{b} = \tilde{\beta} (1-\tilde{h}) 4\pi \tilde{\mu} l_s^2
\tilde{T}_{\rm str}\,.
\end{equation}
Substituting this in Eq.~\eqref{hageab} we get the following
equation for the Hagedorn temperature
\begin{equation}
\tilde T_H=\left[  \sqrt{8\pi} (1-\tilde h) \tilde{T}_{\rm str}
\left({\rm Li}_{\frac{3}{2}}(e^{-2\tilde{h}/\tilde
T_H})\right)^{-1}\right]^{2/3} \label{hagstring}.
\end{equation}
This result is in agreement with the result of
\cite{Harmark:2006ta} for $\tilde{h}=0$. Note that from Section
\ref{sec:decpp} we know that $\tilde{T}_{\rm str}$ is large and since
$0\leq \tilde{h} \leq 1$ we get that $\tilde{h}/\tilde{T}_H \ll
1$. It is therefore sensible to expand the Polylogarithm function
in \eqref{hagstring}.

We can now compare the equation for the Hagedorn temperature
\eqref{hagstring} to the gauge theory side. Using \eqref{redadsdic}
it is easy to see that Eq.~\eqref{hagstring} matches with
Eq.~\eqref{Hagh} on the gauge theory side. Thus, we have
successfully matched the Hagedorn temperature as computed in planar
$\CN=4$ SYM on $\R \times S^3$ in the decoupling limit
\eqref{declim} for $\tilde{\lambda} \gg 1$ with the Hagedorn
temperature computed in free string theory on $\ads_5\times S^5$ in
the decoupling limit given by \eqref{microadslim} and \eqref{epsom}
for $\tilde{T}_{\rm str} \gg 1$.

The fact that we can match the Hagedorn temperature of gauge
theory in the decoupling limit \eqref{declim} and string theory in
the decoupling limit given by \eqref{microadslim} and
\eqref{epsom} is a consequence of the fact that the spectra of the
two theories match in the corresponding decoupling limits, as we
verified in Section \ref{sec:decpp}.

The Hagedorn/deconfinement temperature of planar $\CN=4$ SYM on
$\mathbb{R}\times S^3$ was conjectured to be dual to the Hagedorn
temperature of string theory on AdS$_5\times S^5$ in
\cite{Witten:1998zw,Sundborg:1999ue,Polyakov:2001af,Aharony:2003sx}.
Recently, the first successful matching of the Hagedorn
temperature in AdS/CFT was done in \cite{Harmark:2006ta}. The
above matching of the Hagedorn temperature is an extension of
that.

\subsubsection*{Decoupling limit of the Hagedorn singularity}

As we remarked above, the matching of the gauge theory and string
theory Hagedorn temperature can be seen as a consequence of the
matching of the spectra \eqref{lowspec} and \eqref{spec1}. However,
as a consistency check, we show in the following that the
computation of the Hagedorn temperature for the string theory is
consistent with the decoupling limit \eqref{pplim} that we take on
the pp-wave spectrum \eqref{fullspec1}--\eqref{fullspec2}. We do
that by computing the Hagedorn temperature using the full pp-wave
spectrum \eqref{fullspec1}--\eqref{fullspec2} and then taking the
decoupling limit of the resulting equation for the Hagedorn
singularity.

The starting point is now the partition function \eqref{fab2} and
the computation of the Hagedorn singularity can be seen as a
generalization of the computation in \cite{Sugawara:2003qc} to the
case with an arbitrary parameter $\eta$ in the spectrum
\eqref{fullspec1}--\eqref{fullspec2}. We get the following equation
for the Hagedorn singularity
\begin{equation}
b=4l_s^2\mu \sum_{p=1}^{\infty}\frac{1}{p}\left[3+\cosh(\eta\mu a
p)-4(-1)^p\cosh(\frac{\eta\mu ap}{2})\right]K_1(\mu a
p)\label{hagfull}
\end{equation}
where $K_\nu(x)$ is the modified Bessel function of the second kind.
This equation for the Hagedorn singularity contains as special cases
both the Hagedorn singularity for the $\eta=1$ case corresponding to
the Flat Direction pp-wave background
\eqref{eq:flatdirmet}--\eqref{flatdirF5} which is considered in
\cite{Sugawara:2003qc,Harmark:2006ta} and the $\eta=0$ case
corresponding to the BMN pp-wave background
\eqref{BMNmet}--\eqref{BMNF5} considered in
\cite{PandoZayas:2002hh,Greene:2002cd,Sugawara:2002rs,Brower:2002zx,
Grignani:2003cs}.

The parameters $a$, $b$ and $\eta$ can be expressed in terms of
quantities relating to strings on $\ads_5\times S^5$ as follows
\begin{equation}
a=\frac{\mu \tilde{\beta}}{\tilde{\mu}^2} , \quad
b = \tilde{\beta} (1-\tilde h) 4\pi \mu l_s^2 T_{\rm str} , \quad
\eta=1-2h \label{abh}.
\end{equation}
Now we take the limit \eqref{pplim} of the equation for the Hagedorn
temperature and we use Eq.~\eqref{abh}. It is easy to see that the
only non-vanishing contribution in the limit \eqref{pplim} comes
from the $M_n$ oscillators in the spectrum \eqref{fullspec1} while
all the other terms vanish. This shows that in the decoupling limit
those are precisely the only modes that survive. The result we get
for the Hagedorn temperature is again \eqref{hagstring}.

%%%%%%%%%%%%%%%%%%%%%%%%%%%%%%%%%%%%%%%%%%%%%%%%%%%%%%%%%%%%%%

\section{Discussion and conclusions}
\label{sec:concl}

In this paper we have modified the decoupling limit found in
\cite{Harmark:2006di} to obtain an equivalence between planar
$\CN=4$ SYM on $\R \times S^3$ in the limit \eqref{declim} and the
ferromagnetic $XXX_{1/2}$ Heisenberg spin chain in an external
magnetic field. The difference with the situation considered in
\cite{Harmark:2006di, Harmark:2006ta} is the extra parameter $\tilde
h$ that plays the role of a magnetic field for the Heisenberg chain
and can be regarded as an effective chemical potential for the gauge
theory. The presence of the magnetic field $\tilde h$ breaks the the
degeneracy of the vacuum and leaves a unique vacuum state ${\rm
Tr}(Z^L)$. It furthermore modifies the low energy spectrum in a
non-trivial way such that it cannot be obtained for small
$\tilde{h}$ as a perturbation of the case with zero magnetic field.
This is an effect which is well known for spin systems.

As in the case of zero magnetic field analyzed in
\cite{Harmark:2006di,Harmark:2006ta} only the $SU(2)$ sector
survives the limit \eqref{declim}. Moreover, the Hamiltonian
truncates to $H = D_0 + \tilde{\lambda} D_2 - 2\tilde{h}S_z$ which
means that it consists of terms coming from the bare plus the
one-loop part of the dilation operator. This has the consequence
that we can study the resulting decoupled sector of $\CN=4$ SYM for
any value of the effective coupling $\tilde \lambda$. For small
$\tilde{\lambda}$ we show that the first order term in
$\tilde{\lambda}$ in the effective Hagedorn temperature
$\tilde{T}_H$ comes from a one-loop correction in $\CN=4$ SYM on
$\R\times S^3$. Similarly, the $\tilde{\lambda}^n$ term comes from
an $n$-loop correction. Therefore the large $\tilde{\lambda}$ regime
can be seen as coming from the strong coupling regime of $\CN=4$ SYM
on $\R\times S^3$, even though we have that the 't Hooft coupling
$\lambda$ goes to zero in the decoupling limit \eqref{declim}. The
truncation of the Hamiltonian thus has the consequence that we have
a way to study aspects of the strong coupling regime of the gauge
theory.

Following \cite{Harmark:2006ta} we consider the decoupling limit
\eqref{microadslim} of strings on $\ads_5\times S^5$ which is dual
to the gauge theory decoupling limit \eqref{declim}. We find a
Penrose limit consistent with the decoupling limit
\eqref{microadslim}. This leads us to consider type IIB strings
propagating in the pp-wave background
\eqref{ppwmet}--\eqref{ppwF5}. The extra parameter $h$ on the gauge
theory side coming from the difference between the chemical
potentials emerges as a parameter in the pp-wave background,
signifying an angular velocity in one of the two-planes. This new
parameter allows us to get a pp-wave background that includes as
special cases the BMN pp-wave background
\cite{Blau:2001ne,Berenstein:2002jq} and the Flat Direction
pp-wave background of \cite{Michelson:2002wa,Bertolini:2002nr}.
Indeed, the parameter $h$ measures the departure from the critical
angular velocity giving the Flat Direction pp-wave background.
Having $h>0$ breaks the explicit isometry of the flat direction
and this is analogous to the breaking of the degeneracy of the
vacuum states on the gauge-theory/spin-chain side.

The decoupling limit is implemented for the pp-wave background as
the limit \eqref{pplim}. Taking the decoupling limit of the
pp-wave spectrum \eqref{fullspec1}--\eqref{fullspec2} we find the
same spectrum as for large $\tilde{\lambda}$ and large $L$ on the
gauge theory side. The matching of spectra is one of the main
results of this paper. It is highly non-trivial since we are
matching a spectrum computed for weak 't Hooft coupling on the
gauge theory side to a spectrum in free string theory. What makes
us able to match the spectra is in part our ability to study the
strong coupling regime of the gauge theory by having
$\tilde{\lambda} \gg 1$, as described above. Another important
ingredient is the fact that the pp-wave background is the
maximally supersymmetric background pp-wave of \cite{Blau:2001ne}
which is an exact background of type IIB string theory.

From the matching of the spectra it follows that we can match a
limit of the Hagedorn temperature of string theory on the magnetic
pp-wave background \eqref{ppwmet}--\eqref{ppwF5} to the Hagedorn
temperature of weakly coupled planar $\CN=4$ SYM on $\R \times S^3$
in the limit \eqref{declim} for large $\tilde{\lambda}$ and for any
value of the parameter $\tilde h$. This generalizes the matching of
the Hagedorn temperature for $\tilde{h}=0$ in \cite{Harmark:2006ta}.

In conclusion, we have obtained a triality between planar $\CN=4$
SYM on $\R\times S^3$ in the decoupling limit \eqref{declim}, the
ferromagnetic $XXX_{1/2}$ Heisenberg spin chain coupled to an
external magnetic field and free type IIB string theory on
$\ads_5\times S^5$ in the limit \eqref{microadslim}. The difference
with \cite{Harmark:2006ta} is the extra parameter $\tilde{h}$.
However, as in \cite{Harmark:2006ta}, we have that the Heisenberg
chain with magnetic field is integrable which means that we have
found a solvable sector of AdS/CFT.

One future direction which would be interesting to examine is to
modify the $SU(2|3)$ decoupling limit of $\CN=4$ SYM on $\R \times
S^3$ found in \cite{Harmark:2006di} in a similar way as we
modified the $SU(2)$ limit \eqref{gaugelim} in this paper. The
resulting limit is
\begin{equation}
\label{su23lim} \begin{array}{c} \ds \Omega \rightarrow 1 , \quad
\tilde{T} \equiv \frac{T}{1-\Omega} \ \mbox{fixed} , \quad
\tilde{h}_1 \equiv \frac{h_1}{1-\Omega} \ \mbox{fixed} , \quad
\tilde{h}_2 \equiv \frac{h_2}{1-\Omega} \ \mbox{fixed} , \\[4mm] \ds
\tilde{\lambda} \equiv \frac{\lambda}{1-\Omega} \ \mbox{fixed},
\quad N\ \mbox{fixed},
\end{array}
\end{equation}
with $\Omega \equiv (\Omega_1+\Omega_2+\Omega_3)/3$, $h_1 \equiv
(\Omega_1-\Omega_2)/2$ and $h_2 \equiv (\Omega_2-\Omega_3)/2$. For
$\tilde{h}_1=\tilde{h}_2=0$ this reduces to the $SU(2|3)$ limit of
\cite{Harmark:2006di}. The full partition function of $\CN=4$ SYM on
$\R \times S^3$ in the limit \eqref{su23lim} becomes

\begin{equation}
\label{Zsu23} Z(\tilde{\beta},\tilde{h}_1,\tilde{h}_2) = \tr
\left[ \exp \left( -\tilde{\beta} \left\{ D_0 + \tilde{\lambda}
D_2 - \frac{2}{3} \tilde{h}_1 (2J_1+J_2+J_3) - \frac{2}{3}
\tilde{h}_2 (J_1+J_2+2J_3) \right\} \right) \right]
\end{equation}
where the trace is over the $SU(2|3)$ sector of $\CN=4$ SYM
corresponding to the operators with $D_0=J_1+J_2+J_3$. Here $D_2$ is
an extension of the $D_2$ operator given by Eq.~\eqref{D2} in the
$SU(2)$ sector with the permutation operator now being the graded
permutation operator. The interesting new feature in the $SU(2|3)$
sector is the presence of fermions. In order to find a string theory
dual, it seems evident that one should consider a pp-wave background
with two independent angular rotations in two orthogonal planes.

Another interesting direction that one could pursue is to take a
further decoupling limit in the decoupled sector of $\CN=4$ SYM on
$\R\times S^3$ found in this paper. This is possible since we have
introduced the extra parameter $\tilde{h}$ in the decoupled theory
on the $SU(2)$. A particularly interesting limit is
\begin{equation}
\tilde{\lambda} \rightarrow \infty  , \quad
\tilde{h} \rightarrow 0 , \quad
\tilde{\lambda}(1-\tilde{h})^2 \ \mbox{fixed}.
\end{equation}
In this limit we are left only with the chiral primaries of the
$SU(2)$ sector. However, the Hagedorn temperature $\tilde{T}_H$
remains finite, as is evident from \eqref{Hagh12}. Thus, we have a
phase transition in the supersymmetric sector of $\CN=4$ SYM. This
is similar in spirit to \cite{Silva:2006st}. It would be interesting
to explore this further also on the string theory side, and in
particular to see if there is a connection with supersymmetric AdS
black holes.

Finally, we note that there are several interesting recent works in
the context of weakly coupled thermal $\CN=4$ SYM on $\R\times S^3$
\cite{AlvarezGaume:2006jg,Hollowood:2006xb,Hartnoll:2006pj,Festuccia:2006sa}
and related theories with less supersymmetry
\cite{Karch:2006bv,Hollowood:2006cq,Hikida:2006qb}. We believe that
it could be interesting to combine studies of this kind with
decoupling limits as presented in this paper, since this gives a way
to explore the strongly coupled regime of the gauge theory and to
relate gauge theory computations directly to the dual string theory.

%%%%%%%%%%%%%%%%%%%%%%%%%%%%%%%%%%%%%%%%%%%%%%%%%%%%%%%%%%%%%%

\section*{Acknowledgments}

We thank Jaume Gomis, Gianluca Grignani and Olav Sylju{\aa}sen
for nice and useful discussions. T.H. thanks the Carlsberg Foundation for
support. The work of M.O. is supported in part by the European
Community's Human Potential Programme under contract
MRTN-CT-2004-005104 `Constituents, fundamental forces and
symmetries of the universe'.

\begin{appendix}

%%%%%%%%%%%%%%%%%%%%%%%%%%%%%%%%%%%%%%%%%%%%%%%%%%%%%%%%%%%%%%

\section{Penrose limits}
\label{app:penrose}

In this appendix we write down the two Penrose limits of
$\ads_5\times S^5$ background \eqref{adsmet}--\eqref{adsF5} that we
compare our Penrose limit to in Section \ref{sec:string}. Both
limits give the maximally supersymmetric pp-wave background of
\cite{Blau:2001ne} but in two different coordinate systems. Note
that we use the rescaled AdS radius \eqref{Rtilde} in the limits.

\subsubsection*{BMN Penrose limit}

We write here the {\sl BMN Penrose limit} \cite{Berenstein:2002jq}
(see also \cite{Blau:2002dy}). We define the coordinates
\begin{equation}
x^- = \frac{1}{2} \mu \tilde{R}^2 (t-\chi), \quad x^+ =
\frac{1}{2\mu} (t+\chi), \quad r = \tilde{R} \rho, \quad \tilde{r} =
\tilde{R} \theta , \quad \bar{r} = \tilde{R} \psi.
\end{equation}
The BMN Penrose limit is then
\begin{equation}
\label{BMNlimit} \tilde{R} \rightarrow \infty , \quad
x^+,x^-,r,\Omega_3',\tilde{r},\alpha,\bar{r},\phi \ \mbox{fixed},
\end{equation}
giving the pp-wave background of \cite{Blau:2001ne} with metric
\begin{equation}
\label{BMNmet} \frac{1}{\sqrt{\epsilon}} ds^2 = - 4 dx^+ dx^- -
\mu^2 \sum_{i=1}^8 x^i x^i (dz^+)^2 + \sum_{i=1}^8 dx^i dx^i
\end{equation}
and five-form field strength
\begin{equation}
\label{BMNF5} \frac{1}{\epsilon} F_{(5)} = 2 \,\mu \,dx^+ \left(dx^1
dx^2 dx^3 dx^4 + dx^5 dx^6 dx^7 dx^8 \right).
\end{equation}
We denote this background as the {\sl BMN pp-wave background} since
it is the maximally supersymmetric pp-wave background of
\cite{Blau:2001ne} in the coordinates used in
\cite{Berenstein:2002jq}. The background
\eqref{BMNmet}--\eqref{BMNF5} corresponds to
\eqref{eq:transformedmetric} with $\eta = C = 0$.

%%%%%%%%%%%%%%%%%%%%%%%%%%%%%%%%%%
\subsubsection*{Flat Direction Penrose limit}

We write here the Penrose limit of \cite{Bertolini:2002nr} giving
the pp-wave background of \cite{Michelson:2002wa} corresponding to
the maximally supersymmetric pp-wave of \cite{Blau:2001ne} in a
coordinate system in which we have explicitly a flat direction. This
Penrose limit is denoted the {\sl Flat Direction Penrose limit} in
the main text.

Define $\phi_+$ and $\phi_-$ by
\begin{equation}
\chi = \phi_+ + \phi_-\,, \quad \phi = \phi_+ - \phi_-\, ,
\end{equation}
in terms of which the three-sphere metric \eqref{3sph} is
\begin{equation}
d\Omega_3^2 = d\psi^2 + d\phi_+^2 + d\phi_-^2 + 2\cos(2\psi)
d\phi_+d\phi_- \;.
\end{equation}
We define the coordinates
\begin{equation}
z^- = \frac{1}{2} \mu \tilde{R}^2 (t-\phi_+), \quad
 z^+ =\frac{1}{2\mu} (t+\phi_+),
\end{equation}
\begin{equation}
z^1 = \tilde{R} \phi_-, \quad
z^2 = \tilde{R} \left( \psi -\frac{\pi}{4} \right) , \quad
r=\tilde{R}\rho , \quad
\tilde{r} = \tilde{R} \theta.
\end{equation}
The Flat Direction Penrose limit is then (see also
\cite{Harmark:2006ta})
\begin{equation}
\tilde{R} \rightarrow \infty , \quad
z^+,z^- , z^1 , z^2,r,\Omega_3',\tilde{r},\alpha \ \mbox{fixed}.
\end{equation}
This gives the pp-wave background
\begin{equation}
\label{eq:flatdirmet} \frac{1}{\sqrt{\epsilon}} ds^2 = - 4 dz^+ dz^-
 - \mu^2 \sum_{I=3}^8 (dz^I)^2 (dz^+)^2 + \sum_{i=1}^8  dz^i dz^i -
4\mu z^2 dz^1 dz^+,
\end{equation}
\begin{equation}
\label{flatdirF5} \frac{1}{\epsilon}F_{(5)} = 2 \,\mu \,dz^+
\left(dz^1 dz^2 dz^3 dz^4 + dz^5 dz^6 dz^7 dz^8 \right).
\end{equation}
This background corresponds to \eqref{eq:transformedmetric} with
$\eta = C = 1$.

%%%%%%%%%%%%%%%%%%%%%%%%%%%%%%%%%%%%%%%%%%%%%%%%%%%%%%%%%%%%%%
\section{The magnetic pp-wave}
\label{app:ppwave}

In this appendix we find new pp-wave backgrounds by applying a
time-dependent coordinate transformation to the maximally
supersymmetric pp-wave background of \cite{Blau:2001ne} in the
canonical coordinate system used in
\cite{Blau:2001ne,Berenstein:2002jq}, here denoted as the BMN
pp-wave background. We find a two parameter family of pp-wave
backgrounds that includes as special cases the BMN pp-wave
background \cite{Blau:2001ne,Berenstein:2002jq} and the Flat
Direction pp-wave background of
\cite{Michelson:2002wa,Bertolini:2002nr}. In addition, we get new
{\sl magnetic pp-wave} backgrounds. We call these backgrounds
magnetic because the bosonic Hamiltonian has the same form as the
Hamiltonian of a Newtonian point particle moving in a constant
magnetic field.%
\footnote{See \cite{Gomis:2002km} for a study of a particular
magnetic pp-wave background.}

%%%%%%%%%%%%%%%%%%%%%%%%%%%%%%%%%%%%%%%
\subsection{Coordinate transformation for general $\eta$ and $C$}
\label{sec:coordtrans}

The BMN pp-wave background is given by \eqref{BMNmet}--\eqref{BMNF5}.
We consider here the coordinate transformation
\begin{equation}
\label{eq:coordtrans}
\begin{gathered}
x^- = z^- + \frac{\mu}{2} C z^1 z^2 , \\
\vecto{x^1}{x^2} = \matrto{\cos (\eta \mu z^+) }{-\sin (\eta \mu
z^+) }{\sin (\eta \mu z^+)}{\cos (\eta \mu z^+)} \vecto{z^1}{z^2},
\end{gathered}
\end{equation}
leaving all other coordinates invariant.  The metric \eqref{BMNmet}
becomes
\begin{eqnarray}
\label{eq:transformedmetric} \frac{ds^2}{\sqrt{\epsilon}} &=& - 4
dz^+ dz^- + dz^i dz^i - \mu^2(1-\eta^2)
\Big((z^1)^2+(z^2)^2\Big) (dz^+)^2 - \mu^2 \sum_{I=3}^8 (dz^I)^2 (dz^+)^2 \nn \\
&& - 2\mu (C-\eta) z^1 dz^2 dz^+ - 2\mu (C+\eta) z^2 dz^1dz^+
\end{eqnarray}
while the five-form field \eqref{BMNF5} is
\begin{equation}
\label{eq:transformedF5} \frac{1}{\epsilon}F_{(5)} = 2 \,\mu \,dz^+
\left(dz^1 dz^2 dz^3 dz^4 + dz^5 dz^6 dz^7 dz^8 \right),
\end{equation}
i.e.\ it is invariant under the coordinate transformation.

The parameters $\eta$ and $C$ can a priori take any value. Two
special cases are $\eta = C = 0$, which gives back the BMN pp-wave
background \eqref{BMNmet}--\eqref{BMNF5}, and $\eta = C =1$, which
gives the Flat Direction pp-wave background
\eqref{eq:flatdirmet}--\eqref{flatdirF5}. We will also be interested
in the case with $0 < \eta <1$ and  $C=0$. This gives the new
pp-wave backgrounds that we find in Section \ref{sec:string} to be
dual to the Heisenberg spin-chain in an external magnetic field,
when taking a limit with $\eta \rightarrow 1$.

%%%%%%%%%%%%%%%%%%%%%%%%%%%%%%%%%%%%%%%%%%%
\subsection{String theory spectrum}
\label{sec:fullspec}

In this section we derive the Hamiltonian and the spectrum of string
theory on the magnetic pp-wave background
\eqref{eq:transformedmetric}--\eqref{eq:transformedF5}. This is only
a minor generalization of Section 3 in Ref.~\cite{Bertolini:2002nr}
and we will therefore be brief.

In the light-cone gauge, $Z^+ = l_s^2 p^+ \tau$, the light-cone
Lagrangian
 of the bosonic sigma-model is
\begin{equation}
\begin{array}{c} \ds
 \frac{l_s^2p^+}{\sqrt{\epsilon}} \mathcal{L}_{\rm lc}
    = - \frac{1}{4\pi l_s^2} \left( \partial^\alpha Z^i \partial_\alpha Z^i
    + f^2 (1-\eta^2) \Big((Z^1)^2+(Z^2)^2\Big)
      + f^2 \sum_{I=3}^8 (Z^I)^2 \right. \\[5mm] \ds \left.
    +2f (C-\eta) Z^1 \dot{Z}^2
      + 2f (C+\eta) Z^2 \dot{Z}^1 \right)
\end{array}
\end{equation}
where we have defined $f = \mu l_s^2 p^+$. The conjugate momenta are
\begin{equation}
\label{eq:conjugatemom}
\Pi_1 = \frac{\dot{Z}^1 - f (C+\eta) Z^2}{2\pi l_s^2}, \quad
\Pi_2 = \frac{\dot{Z}^2 - f (C-\eta) Z^1 }{2\pi l_s^2}, \quad
\Pi_I = \frac{\dot{Z}^I}{2\pi l_s^2},
\end{equation}
giving the bosonic Hamiltonian
\begin{equation}
\label{eq:hamiltonian} \frac{l_s^2p^+}{\sqrt{\epsilon}}H^B_{\rm lc}
= \frac{1}{4\pi l_s^2} \int_0^{2\pi} d\sigma \left[ \dot{Z}^i
\dot{Z}^i + (Z^i)' (Z^i)' + f^2 (1-\eta^2) \big( (Z^1)^2 + (Z^2)^2
\big) + f^2 \sum_{I=3}^8 (Z^I)^2 \right].
\end{equation}
Note that the parameter $C$ has dropped out of the Hamiltonian. That
is because we have expressed it in terms of the velocities and not
the true hamiltonian variables which are the conjugate momenta.

{}From the Lagrangian we find the equations of motion, expand the
solutions in oscillator modes, and quantize in the canonical way.
The bosonic Hamiltonian can then be written in terms of number
operators as
\begin{align}
\frac{l_s^2p^+}{\sqrt{\epsilon}}H^\textrm{B}_\textrm{lc} = \sum_{n =
-\infty}^{\infty} \left( (\omega_n + \eta f) N_n + (\omega_n - \eta
f) M_n + \sum_{I=3}^{8} \omega_n N_n^I \right)
\end{align}
where
\begin{align}
\omega_n = \sqrt{n^2 +f^2}, \quad \textrm{for all } n \in \Z.
\end{align}

If we consider the range of $\eta$, we find that for $\eta > 1$ the
mode $M_0$ has the spectrum
\begin{equation}
f(1-\eta)M_0
\end{equation}
and therefore the Hamiltonian $H_{\rm lc}^{\rm B}$ can have
arbitrarily large negative energies, signaling an instability. This
suggests that having $\eta > 1$ is not possible, and that $\eta=1$
is a critical value for  $\eta$. Similarly, from the $N_0$ mode we
get the condition $\eta \geq -1$.  Thus, the physically acceptable
range of $\eta$ is
\begin{equation}
-1 \leq \eta \leq 1.
\end{equation}

We find the fermionic part of the spectrum in exactly the same way
as was done in Section 3.2 of \cite{Bertolini:2002nr}. Starting with
$\theta^{A}$ as a Majorana-Weyl spinor with 16 non-vanishing
components and $A=1,2$, we choose the light-cone gauge
\begin{equation}\label{fermlcc}
    Z^+ = l_s^2\ p^+ \tau, \qquad \Gamma^{\hat+} \theta^A=0.
\end{equation}
The Green-Schwarz fermionic action is then given by
\cite{Metsaev:2002re}
\begin{equation}\label{fermact}
    \frac{l_s^2p^+}{\sqrt{\epsilon}} S^{\text{F}}_{\text{lc}} = \frac{i}{4\pi l_s^2} \int d\tau d\sigma
        \left[ \left( \eta^{\alpha\beta} \delta_{AB} -
        \epsilon^{\alpha\beta}(\sigma_ 3)_{AB} \right) \partial_\alpha
        Z^+ \bar{\theta}^A \Gamma_+ ( \mathcal{D}_\beta \theta )^B
        \right]
\end{equation}
and the generalized covariant derivative takes the form
\cite{Metsaev:2002re}
\begin{equation}\label{fermD}
    \mathcal{D}_\beta = \partial_\beta +\frac{1}{4} \partial_\beta
        Z^+ \left( \omega_{+\hat\rho\hat\sigma} \Gamma^{\hat\rho\hat\sigma}
        -\frac{1}{2\cdot 5!} F_{\lambda\nu\rho\sigma\kappa}
        \Gamma^{\lambda\nu\rho\sigma\kappa} i\sigma_2 \Gamma_+
        \right).
\end{equation}

In order to proceed, we need to find all the relevant components of
the spin connection $\omega_{\mu\hat a\hat b}$. We put a hat on flat
indices in the tangent space to distinguish them from the curved
indices in the spacetime. For our purposes here it is enough to find
the components where the curved index is $+$. It turns out that the
only relevant components are
\begin{align}
\label{eq:relevantspcon} \omega_{+\hat1\hat2} = -\eta\mu, \quad
\omega_{+\hat2\hat1} = \eta\mu.
\end{align}
The other components either vanish, like $\omega_{+\hat I\hat J} =
0$, or are contracted with $\Gamma^{\hat +}$ in the covariant
derivative and are thus killed in the light-cone gauge, like
$\omega_{+\hat + \hat I} = -\mu^2z^I$.

Following the same steps as in the paper \cite{Bertolini:2002nr}
(and using the same notation), we arrive at the fermionic light-cone
action
\begin{align}
\frac{l_s^2p^+}{\sqrt{\epsilon}} S^\textrm{F}_\textrm{lc} =
\frac{i}{2\pi}\int d\tau d\sigma \left[ S^1 \left(\partial_+
-\frac{\eta f}{2}\gamma^{12} \right)S^1 + S^2 \left(\partial_-
-\frac{\eta f}{2}\gamma^{12} \right)S^2 - 2fS^1\Pi S^2 \right]
\end{align}
where $\partial_{\pm} = \partial_\tau \pm \partial_{\sigma}\,$ and
$\Pi=\gamma^{1234}$. Note that there are two ``sources" of terms
involving $f= \mu l_s^2 p^+$. The $S^A\eta f\gamma^{12}S^A$ terms
come from the spin connection \eqref{eq:relevantspcon} and therefore
the $f$ in \cite{Bertolini:2002nr} is replaced by $\eta f$ here,
while the $2fS^1\Pi S^2$ term comes from the five-form field and
thus does not contain any $\eta$.

Again, following the same steps as in Sections 3.2--3.3 of
\cite{Bertolini:2002nr}, we find that the fermionic part of the
Hamiltonian is given by
\begin{align}
\frac{l_s^2p^+}{\sqrt{\epsilon}}H_\textrm{lc}^\textrm{F} &= \sum_{n
= -\infty}^{+\infty}
    S^\dagger_n \left(\omega_n + i\frac{\eta f}{2}\gamma^{12}\right)S_n\\
&= \sum_{n = -\infty}^{+\infty} \left[ \sum_{b=1}^{4}\left( \omega_n
-\frac{\eta f}{2}\right) F_n^{(b)} + \sum_{b=5}^{8}\left( \omega_n
+\frac{\eta f}{2}\right) F_n^{(b)} \right].
\end{align}
The full Hamiltonian of quantized strings on the magnetic pp-wave in
light-cone gauge is therefore
\begin{align}
\label{fullspec1} \frac{l_s^2p^+}{\sqrt{\epsilon}}H_\textrm{lc} &=
\sum_{n = -\infty}^{+\infty} \left( (\omega_n + \eta f) N_n +
(\omega_n - \eta f) M_n
+ \sum_{I=3}^{8} \omega_n N_n^I \right) \nonumber \\
&+ \sum_{n = -\infty}^{+\infty} \left( \sum_{b=1}^{4}\left( \omega_n
-\frac{\eta f}{2}\right) F_n^{(b)} + \sum_{b=5}^{8}\left( \omega_n
+\frac{\eta f}{2}\right) F_n^{(b)} \right)
\end{align}
with the level matching condition
\begin{align}
\label{fullspec2} \sum_{n = -\infty}^{+\infty} n \left( N_n + M_n +
\sum_{I=3}^{8}  N_n^I + \sum_{b=1}^{8} F_n^{(b)} \right) = 0.
\end{align}

%%%%%%%%%%%%%%%%%%%%%%%%%%%%%%%%%%%%%%%%%%
\subsection{Physical interpretation}
\label{app:physint}

We can compare the bosonic Hamiltonian of the magnetic pp-wave in
equation \eqref{eq:hamiltonian} to Newtonian physics of a charged
particle moving in a constant magnetic field.  We will see that the
parameter $\eta$ corresponds to the strength of the magnetic field
while $C$ is a gauge choice.

To be more precise, let the particle move in the $(z^1,z^2)$ plane
with a constant magnetic field $\vec{B} = B \vec{e}_3$ perpendicular
to the plane. The particle is furthermore connected to the origin
with a spring of spring constant $k$. We can take the vector
potential to be
\begin{equation}
\vec A = \frac{1}{2} \vec B \times \vec z = \frac{B}{2} \left(-z^2,
z^1, 0 \right)^T
\end{equation}
but we are also free to add any $\vec \nabla \phi$ to the vector
potential since that only amounts to choosing a different gauge.
Let's take $\phi = \gamma z^1 z^2$ so that
\begin{equation}
\vec \nabla \phi = \gamma \left(z^2,  z^1,  0\right)^T.
\end{equation}

To find the Hamiltonian of this system, we start with the
Hamiltonian of an uncharged particle and simply replace $\vec p$
with $\vec p + \vec A$. This gives
\begin{align}
H &= \frac{1}{2m} (\vec p + \vec A)^2 + \frac{1}{2}k \vec z^2 \\
\label{eq:NewtonHamilton} &= \frac{1}{2m} \left\{
         \left( p_1 + (\gamma - B/2)z^2\right)^2
      + \left( p_2 + (\gamma + B/2)z^1\right)^2
    \right\}
+ \frac{1}{2} k \left( (z^1)^2 + (z^2)^2 \right).
\end{align}
This Hamiltonian should be compared to the $Z^1$ and $Z^2$ part of
the bosonic Hamiltonian \eqref{eq:hamiltonian} with the velocities
$\dot Z^i$ replaced by the conjugate momenta $\Pi_i$ from
Eq.~\eqref{eq:conjugatemom}.  Let's set $2\pi \ell_s^2 = m = 1$ for
now. The relevant part of the bosonic Hamiltonian is then
\begin{align}
\frac{1}{2} \int_0^{2\pi} d\sigma \left[ \left(\Pi_1 + f(C+\eta)Z^2
\right)^2 + \left(\Pi_2 + f(C-\eta)Z^1 \right)^2 + f^2 (1-\eta^2)
\big( (Z^1)^2 + (Z^2)^2 \big) \right].
\end{align}
Comparing this to Eq.~\eqref{eq:NewtonHamilton} we find the
following dictionary
\begin{align}
-\frac{B}{2}  \leftrightarrow   f \eta, \qquad \gamma
\leftrightarrow  f C, \qquad k  \leftrightarrow   f^2(1-\eta^2).
\end{align}
This shows that the parameter $\eta$ corresponds to the strength of
the magnetic field and that $C$ is a gauge choice for the vector
potential.

The BMN background corresponds to having no magnetic field while in
the Flat Direction background, the gauge $C=\eta$ was chosen to
eliminate all dependence of $Z^1$ from the Hamiltonian.

%%%%%%%%%%%%%%%%%%%%%%%%%%%%%%%%%%%%%%%%%%%%%%%%%%%%%%%%%%

\end{appendix}

%%%%%%%%%%%%%%%%%%%%%%%%%%%%%%%%%%%%%%%%%%%%%%%%%%%%%%%%%%

\providecommand{\href}[2]{#2}\begingroup\raggedright\endgroup


\begin{thebibliography}{10}

\bibitem{Maldacena:1997re}
J.~Maldacena, ``The large {N} limit of superconformal field theories
and
  supergravity,'' {\em Adv. Theor. Math. Phys.} {\bf 2} (1998) 231--252,
\href{http://www.arXiv.org/abs/hep-th/9711200}{{\tt
hep-th/9711200}}.
%%CITATION = 00203,2,231;%%.

\bibitem{Gubser:1998bc}
S.~S. Gubser, I.~R. Klebanov, and A.~M. Polyakov, ``Gauge theory
correlators
  from noncritical string theory,'' {\em Phys. Lett.} {\bf B428} (1998) 105,
\href{http://www.arXiv.org/abs/hep-th/9802109}{{\tt
hep-th/9802109}}.
%%CITATION = PHLTA,B428,105;%%.

\bibitem{Witten:1998qj}
E.~Witten, ``{Anti-de Sitter} space and holography,'' {\em Adv.
Theor. Math.
  Phys.} {\bf 2} (1998) 253,
\href{http://www.arXiv.org/abs/hep-th/9802150}{{\tt
hep-th/9802150}}.
%%CITATION = 00203,2,253;%%.

\bibitem{Harmark:2006di}
T.~Harmark and M.~Orselli, ``Quantum mechanical sectors in thermal
{$\CN = 4$}
  super {Yang-Mills} on {$\R \times S^3$},'' {\em Nucl. Phys.} {\bf B757}
  (2006) 117--145,
\href{http://www.arXiv.org/abs/hep-th/0605234}{{\tt
hep-th/0605234}}.
%%CITATION = HEP-TH 0605234;%%.

\bibitem{Harmark:2006ta}
T.~Harmark and M.~Orselli, ``Matching the {Hagedorn} temperature in
  {AdS/CFT},'' {\em Phys. Rev.} {\bf D74} (2006) 126009,
\href{http://www.arXiv.org/abs/hep-th/0608115}{{\tt hep-th/0608115}}.
%%CITATION = HEP-TH 0608115;%%.

\bibitem{Witten:1998zw}
E.~Witten, ``{Anti-de Sitter space}, thermal phase transition, and
confinement
  in gauge theories,'' {\em Adv. Theor. Math. Phys.} {\bf 2} (1998) 505,
\href{http://www.arXiv.org/abs/hep-th/9803131}{{\tt
hep-th/9803131}}.
%%CITATION = 00203,2,505;%%.

\bibitem{Sundborg:1999ue}
B.~Sundborg, ``The {Hagedorn} transition, deconfinement and {$\CN =
4$ SYM}
  theory,'' {\em Nucl. Phys.} {\bf B573} (2000) 349--363,
\href{http://www.arXiv.org/abs/hep-th/9908001}{{\tt
hep-th/9908001}}.
%%CITATION = HEP-TH 9908001;%%.

\bibitem{Polyakov:2001af}
A.~M. Polyakov, ``Gauge fields and space-time,'' {\em Int. J. Mod.
Phys.} {\bf
  A17S1} (2002) 119--136,
\href{http://www.arXiv.org/abs/hep-th/0110196}{{\tt
hep-th/0110196}}.
%%CITATION = HEP-TH 0110196;%%.

\bibitem{Aharony:2003sx}
O.~Aharony, J.~Marsano, S.~Minwalla, K.~Papadodimas, and
M.~Van~Raamsdonk,
  ``The {Hagedorn}/deconfinement phase transition in weakly coupled large {N}
  gauge theories,''
\href{http://www.arXiv.org/abs/hep-th/0310285}{{\tt
hep-th/0310285}}.
%%CITATION = HEP-TH 0310285;%%.

\bibitem{Blau:2001ne}
M.~Blau, J.~Figueroa-O'Farrill, C.~Hull, and G.~Papadopoulos, ``A
new maximally
  supersymmetric background of {IIB} superstring theory,'' {\em JHEP} {\bf 01}
  (2002) 047,
\href{http://www.arXiv.org/abs/hep-th/0110242}{{\tt
hep-th/0110242}}.
%%CITATION = HEP-TH 0110242;%%.

\bibitem{Berenstein:2002jq}
D.~Berenstein, J.~M. Maldacena, and H.~Nastase, ``Strings in flat
space and pp
  waves from {$\CN = 4$} super {Yang Mills},'' {\em JHEP} {\bf 04} (2002) 013,
\href{http://www.arXiv.org/abs/hep-th/0202021}{{\tt
hep-th/0202021}}.
%%CITATION = HEP-TH 0202021;%%.

\bibitem{Michelson:2002wa}
J.~Michelson, ``(twisted) toroidal compactification of pp-waves,''
{\em Phys.
  Rev.} {\bf D66} (2002) 066002,
\href{http://www.arXiv.org/abs/hep-th/0203140}{{\tt
hep-th/0203140}}.
%%CITATION = HEP-TH 0203140;%%.

\bibitem{Bertolini:2002nr}
M.~Bertolini, J.~de~Boer, T.~Harmark, E.~Imeroni, and N.~A. Obers,
``Gauge
  theory description of compactified pp-waves,'' {\em JHEP} {\bf 01} (2003)
  016,
\href{http://www.arXiv.org/abs/hep-th/0209201}{{\tt
hep-th/0209201}}.
%%CITATION = HEP-TH 0209201;%%.

\bibitem{Beisert:2003tq}
N.~Beisert, C.~Kristjansen, and M.~Staudacher, ``The dilatation
operator of
  {$\CN = 4$} super {Yang-Mills} theory,'' {\em Nucl. Phys.} {\bf B664} (2003)
  131--184,
\href{http://www.arXiv.org/abs/hep-th/0303060}{{\tt
hep-th/0303060}}.
%%CITATION = HEP-TH 0303060;%%.

\bibitem{Beisert:2004ry}
N.~Beisert, ``The dilatation operator of {$\CN = 4$} super
{Yang-Mills} theory
  and integrability,'' {\em Phys. Rept.} {\bf 405} (2005) 1--202,
\href{http://www.arXiv.org/abs/hep-th/0407277}{{\tt
hep-th/0407277}}.
%%CITATION = HEP-TH 0407277;%%.

\bibitem{Minahan:2002ve}
J.~A. Minahan and K.~Zarembo, ``The {Bethe-ansatz} for {$\CN = 4$}
super
  {Yang-Mills},'' {\em JHEP} {\bf 03} (2003) 013,
\href{http://www.arXiv.org/abs/hep-th/0212208}{{\tt
hep-th/0212208}}.
%%CITATION = HEP-TH 0212208;%%.

\bibitem{Spradlin:2004pp}
M.~Spradlin and A.~Volovich, ``A pendant for {Polya}: The one-loop
partition
  function of {$\CN = 4$ SYM on $\R \times S^3$},'' {\em Nucl. Phys.} {\bf
  B711} (2005) 199--230,
\href{http://www.arXiv.org/abs/hep-th/0408178}{{\tt
hep-th/0408178}}.
%%CITATION = HEP-TH 0408178;%%.

\bibitem{Yamada:2006rx}
D.~Yamada and L.~G. Yaffe, ``Phase diagram of {$\CN = 4$}
{super-Yang-Mills}
  theory with {R-symmetry} chemical potentials,'' {\em JHEP} {\bf 09} (2006)
  027,
\href{http://www.arXiv.org/abs/hep-th/0602074}{{\tt
hep-th/0602074}}.
%%CITATION = HEP-TH 0602074;%%.

\bibitem{takahashi}
M.~Takahashi, {\em Thermodynamics of One-Dimensional Solvable
Models}.
\newblock Cambridge University Press, 1999.



\bibitem{Takahashi:2000}
M.~Takahashi, ``Simplification of thermodynamic {B}ethe-ansatz equations,''
  \href{http://www.arXiv.org/abs/cond-mat/0010486}{{\tt cond-mat/0010486}}.

\bibitem{Takahashi:2001}
M.~Takahashi, M.~Shiroishi, and A.~Kl{\"u}mper, ``Equivalence of TBA and QTM,''
  {\em J. Phys. A: Math. Gen.} {\bf 34} (2001) L187--L194.



\bibitem{PhysRevLett.89.117201}
M.~Shiroishi and M.~Takahashi, ``Integral equation generates
high-temperature
  expansion of the {Heisenberg} chain,'' {\em Phys. Rev. Lett.} {\bf 89}
  (2002), no.~11, 117201.

\bibitem{Harmark:1999xt}
T.~Harmark and N.~A. Obers, ``Thermodynamics of spinning branes and
their dual
  field theories,'' {\em JHEP} {\bf 01} (2000) 008,
\href{http://www.arXiv.org/abs/hep-th/9910036}{{\tt
hep-th/9910036}}.
%%CITATION = JHEPA,0001,008;%%.

\bibitem{PandoZayas:2002hh}
L.~A. Pando~Zayas and D.~Vaman, ``Strings in {RR} plane wave
background at
  finite temperature,'' {\em Phys. Rev.} {\bf D67} (2003) 106006,
\href{http://www.arXiv.org/abs/hep-th/0208066}{{\tt
hep-th/0208066}}.
%%CITATION = HEP-TH 0208066;%%.

\bibitem{Greene:2002cd}
B.~R. Greene, K.~Schalm, and G.~Shiu, ``On the {Hagedorn} behaviour
of
  {pp-wave} strings and {$\CN = 4$} {SYM} theory at finite {R-charge}
  density,'' {\em Nucl. Phys.} {\bf B652} (2003) 105--126,
\href{http://www.arXiv.org/abs/hep-th/0208163}{{\tt
hep-th/0208163}}.
%%CITATION = HEP-TH 0208163;%%.

\bibitem{Sugawara:2002rs}
Y.~Sugawara, ``Thermal amplitudes in {DLCQ} superstrings on
{pp-waves},'' {\em
  Nucl. Phys.} {\bf B650} (2003) 75--113,
\href{http://www.arXiv.org/abs/hep-th/0209145}{{\tt
hep-th/0209145}}.
%%CITATION = HEP-TH 0209145;%%.

\bibitem{Brower:2002zx}
R.~C. Brower, D.~A. Lowe, and C.-I. Tan, ``Hagedorn transition for
strings on
  {pp-waves} and tori with chemical potentials,'' {\em Nucl. Phys.} {\bf B652}
  (2003) 127--141,
\href{http://www.arXiv.org/abs/hep-th/0211201}{{\tt
hep-th/0211201}}.
%%CITATION = HEP-TH 0211201;%%.

\bibitem{Sugawara:2003qc}
Y.~Sugawara, ``Thermal partition function of superstring on
compactified
  {pp-wave},'' {\em Nucl. Phys.} {\bf B661} (2003) 191--208,
\href{http://www.arXiv.org/abs/hep-th/0301035}{{\tt
hep-th/0301035}}.
%%CITATION = HEP-TH 0301035;%%.

\bibitem{Grignani:2003cs}
G.~Grignani, M.~Orselli, G.~W. Semenoff, and D.~Trancanelli, ``The
superstring
  {Hagedorn} temperature in a {pp-wave} background,'' {\em JHEP} {\bf 06}
  (2003) 006,
\href{http://www.arXiv.org/abs/hep-th/0301186}{{\tt
hep-th/0301186}}.
%%CITATION = HEP-TH 0301186;%%.

\bibitem{Hyun:2003ks}
S.-j. Hyun, J.-D. Park, and S.-H. Yi, ``Thermodynamic behavior of
{IIA} string
  theory on a {pp-wave},'' {\em JHEP} {\bf 11} (2003) 006,
\href{http://www.arXiv.org/abs/hep-th/0304239}{{\tt
hep-th/0304239}}.
%%CITATION = HEP-TH 0304239;%%.

\bibitem{Bigazzi:2003jk}
F.~Bigazzi and A.~L. Cotrone, ``On zero-point energy, stability and
{Hagedorn}
  behavior of type {IIB} strings on {pp-waves},'' {\em JHEP} {\bf 08} (2003)
  052,
\href{http://www.arXiv.org/abs/hep-th/0306102}{{\tt
hep-th/0306102}}.
%%CITATION = HEP-TH 0306102;%%.





\bibitem{Deo:1989bv}
N.~Deo, S.~Jain, and C.-I. Tan, ``String statistical mechanics above
Hagedorn
  energy density,'' {\em Phys. Rev.} {\bf D40} (1989)
2626.
%%CITATION = PHRVA,D40,2626;%%.
%
%\bibitem{Brower:1998an}
R.~C. Brower, J.~McGreevy, and C.~I. Tan, ``Stringy model for QCD at
finite
  density and generalized Hagedorn temperature,''
\href{http://www.arXiv.org/abs/hep-ph/9907258}{{\tt
hep-ph/9907258}}.
%%CITATION = HEP-PH/9907258;%%.
%
%\bibitem{Grignani:2001hb}
G.~Grignani, M.~Orselli, and G.~W. Semenoff, ``Matrix strings in a
B-field,''
  {\em JHEP} {\bf 07} (2001) 004,
\href{http://www.arXiv.org/abs/hep-th/0104112}{{\tt
hep-th/0104112}}.
%%CITATION = HEP-TH/0104112;%%.
%
%\bibitem{Grignani:2001ik}
G.~Grignani, M.~Orselli, and G.~W. Semenoff, ``The target space
dependence of
  the Hagedorn temperature,'' {\em JHEP} {\bf 11} (2001) 058,
\href{http://www.arXiv.org/abs/hep-th/0110152}{{\tt
hep-th/0110152}}.
%%CITATION = HEP-TH/0110152;%%.
%
%\bibitem{DeRisi:2002gt}
G.~De~Risi, G.~Grignani, and M.~Orselli, ``Space / time
noncommutativity in
  string theories without background electric field,'' {\em JHEP} {\bf 12}
  (2002) 031,
\href{http://www.arXiv.org/abs/hep-th/0211056}{{\tt
hep-th/0211056}}.
%%CITATION = HEP-TH/0211056;%%.





\bibitem{Silva:2006st}
P.~J. Silva, ``Phase transitions and statistical mechanics for {BPS} black
  holes in {AdS/CFT},'' {\em JHEP} {\bf 03} (2007) 015,
\href{http://www.arXiv.org/abs/hep-th/0610163}{{\tt hep-th/0610163}}.
%%CITATION = HEP-TH/0610163;%%.

\bibitem{AlvarezGaume:2006jg}
L.~Alvarez-Gaume, P.~Basu, M.~Marino, and S.~R. Wadia, ``Blackhole / string
  transition for the small {Schwarzschild} blackhole of {$\mbox{AdS}_5 \times
  S^5$} and critical unitary matrix models,'' {\em Eur. Phys. J.} {\bf C48}
  (2006) 647--665,
\href{http://www.arXiv.org/abs/hep-th/0605041}{{\tt hep-th/0605041}}.
%%CITATION = HEP-TH/0605041;%%.

\bibitem{Hollowood:2006xb}
T.~Hollowood, S.~P. Kumar, and A.~Naqvi, ``Instabilities of the small black
  hole: A view from {$\CN = 4$} {SYM},'' {\em JHEP} {\bf 01} (2007) 001,
\href{http://www.arXiv.org/abs/hep-th/0607111}{{\tt hep-th/0607111}}.
%%CITATION = HEP-TH/0607111;%%.

\bibitem{Hartnoll:2006pj}
S.~A. Hartnoll and S.~P. Kumar, ``Thermal {$\CN = 4$} {SYM} theory as a {2D}
  {Coulomb} gas,'' {\em Phys. Rev.} {\bf D76} (2007) 026005,
\href{http://www.arXiv.org/abs/hep-th/0610103}{{\tt hep-th/0610103}}.
%%CITATION = HEP-TH/0610103;%%.

\bibitem{Festuccia:2006sa}
G.~Festuccia and H.~Liu, ``The arrow of time, black holes, and quantum mixing
  of large {N} {Yang-Mills} theories,''
\href{http://www.arXiv.org/abs/hep-th/0611098}{{\tt hep-th/0611098}}.
%%CITATION = HEP-TH 0611098;%%.

\bibitem{Karch:2006bv}
A.~Karch and A.~O'Bannon, ``Chiral transition of {$\CN = 4$} super {Yang-Mills}
  with flavor on a 3-sphere,'' {\em Phys. Rev.} {\bf D74} (2006) 085033,
\href{http://www.arXiv.org/abs/hep-th/0605120}{{\tt hep-th/0605120}}.
%%CITATION = HEP-TH/0605120;%%.

\bibitem{Hollowood:2006cq}
T.~J. Hollowood and A.~Naqvi, ``Phase transitions of orientifold gauge theories
  at large {N} in finite volume,'' {\em JHEP} {\bf 04} (2007) 087,
\href{http://www.arXiv.org/abs/hep-th/0609203}{{\tt hep-th/0609203}}.
%%CITATION = HEP-TH/0609203;%%.

\bibitem{Hikida:2006qb}
Y.~Hikida, ``Phase transitions of large {N} orbifold gauge theories,'' {\em
  JHEP} {\bf 12} (2006) 042,
\href{http://www.arXiv.org/abs/hep-th/0610119}{{\tt hep-th/0610119}}.
%%CITATION = HEP-TH/0610119;%%.


\bibitem{Blau:2002dy}
M.~Blau, J.~Figueroa-O'Farrill, C.~Hull, and G.~Papadopoulos,
``{Penrose}
  limits and maximal supersymmetry,'' {\em Class. Quant. Grav.} {\bf 19} (2002)
  L87--L95,
\href{http://arXiv.org/abs/hep-th/0201081}{{\tt hep-th/0201081}}.
%%CITATION = HEP-TH 0201081;%%.

\bibitem{Gomis:2002km}
J.~Gomis and H.~Ooguri, ``Penrose limit of {$\CN = 1$} gauge
theories,'' {\em
  Nucl. Phys.} {\bf B635} (2002) 106--126,
\href{http://www.arXiv.org/abs/hep-th/0202157}{{\tt
hep-th/0202157}}.
%%CITATION = HEP-TH 0202157;%%.

\bibitem{Metsaev:2002re}
R.~R. Metsaev and A.~A. Tseytlin, ``Exactly solvable model of
superstring in
  plane wave {Ramond-Ramond} background,'' {\em Phys. Rev.} {\bf D65} (2002)
  126004,
\href{http://arXiv.org/abs/hep-th/0202109}{{\tt hep-th/0202109}}.
%%CITATION = HEP-TH 0202109;%%.

\end{thebibliography}
\end{document}